\documentclass[sigconf, authorversion, nonacm]{acmart}
\AtBeginDocument{%
  }

\setcopyright{none} 
\copyrightyear{2025}
\acmYear{2025}
\acmDOI{XXXXXXX.XXXXXXX}

\acmConference[MICRO 2025]{The 58th IEEE/ACM International Symposium on Microarchitecture}{October 18--22, 2025}{Seoul, Korea}
\acmISBN{978-X-XXXX-XXXX-X/XX/XX}

\newcommand{\todo}[1]{{\color{red}[#1]}}

\newcommand{\myparagraph}[1]{\vspace{1pt}\noindent\textbf{#1.\xspace}}

\newcommand*{\rom}[1]{\uppercase\expandafter{\romannumeral #1\relax}}

\definecolor{dkgreen}{rgb}{0,0.6,0}
\definecolor{gray}{rgb}{0.5,0.5,0.5}
\definecolor{mauve}{rgb}{0.58,0,0.82}

\newcommand{\system}{Chameleon}

\usepackage[ruled,vlined,noend]{algorithm2e}
\usepackage[]{hyperref}
\DeclareUnicodeCharacter{03BC}{\ensuremath{\mu}}

\usepackage{tikz}

\usepackage{amsmath}
\usepackage{enumitem}
\usepackage{soul}
\usepackage{mydefs,acmart-taps}

\copyrightyear{2025}
\acmYear{2025}
\setcopyright{cc}
\setcctype{by}
\acmConference[MICRO '25]{58th IEEE/ACM International Symposium on
Microarchitecture}{October 18--22, 2025}{Seoul, Republic of Korea}
\acmBooktitle{58th IEEE/ACM International Symposium on Microarchitecture
(MICRO '25), October 18--22, 2025, Seoul, Republic of
Korea}\acmDOI{10.1145/3725843.3756083}
\acmISBN{979-8-4007-1573-0/2025/10}
\begin{document}

\title{\system{}: Adaptive Caching and Scheduling for Many-Adapter LLM Inference Environments}
%\subtitle{\normalsize{MICRO 2025 Submission
%    \textbf{\#673} -- Confidential Draft -- Do NOT Distribute!!}\vspace{-2mm}}
%\author{\normalsize{ISCA 2025 Submission
 %   \textbf{\#NaN} -- Confidential Draft -- Do NOT Distribute!!}}

\author{Nikoleta Iliakopoulou}
\affiliation{
    \institution{University of Illinois at Urbana-Champaign}
    \country{Urbana, USA}
}
\email{nmi4@illinois.edu}

\author{Jovan Stojkovic}
\affiliation{
    \institution{University of Illinois at Urbana-Champaign}
    \country{Urbana, USA}
}
\email{jovans2@illinois.edu}

\author{Chloe Alverti}
\affiliation{
    \institution{University of Illinois at Urbana-Champaign}
    \country{Urbana, USA}
}
\email{xalverti@illinois.edu}

\author{Tianyin Xu}
\affiliation{
    \institution{University of Illinois at Urbana-Champaign}
    \country{Urbana, USA}
}
\email{tyxu@illinois.edu}

\author{Hubertus Franke}
\affiliation{
    \institution{IBM Research}
    \country{Yorktown Heights, USA}
}
\email{frankeh@us.ibm.com}

\author{Josep Torrellas}
\affiliation{
    \institution{University of Illinois at Urbana-Champaign}
    \country{Urbana, USA}
}
\email{torrella@illinois.edu}

\begin{comment}

\author{
Nikoleta Iliakopoulou,
Jovan Stojkovic, 
Chloe Alverti,
Tianyin Xu,
Hubertus Franke$^*$,  
Josep Torrellas
\\
\hspace{-0.9cm}\qquad  University of Illinois Urbana-Champaign 
\qquad  $^*$IBM Research\\
 \{nmi4,jovans2,xalverti,tyxu,torrella\}@illinois.edu, frankeh@us.ibm.com
}
\end{comment}
\renewcommand{\shortauthors}{Nikoleta Iliakopoulou et al.}


%%%%%% -- PAPER CONTENT STARTS-- %%%%%%%%

\begin{abstract}
The effectiveness of LLMs has triggered an exponential rise in their deployment,
imposing substantial demands on inference clusters.
Such clusters often handle numerous concurrent queries for different LLM downstream tasks.  
To handle multi-task settings with vast LLM parameter counts,  Low-Rank Adaptation (LoRA) enables task-specific fine-tuning while sharing most of the base LLM model across tasks.
Hence, it supports concurrent task serving with reduced memory requirements. 
However, existing 
designs
face inefficiencies: they overlook workload heterogeneity, impose high CPU-GPU link bandwidth from frequent adapter loading, and suffer from head-of-line blocking in their schedulers.

To address these challenges,
we present \emph{Chameleon}, a novel LLM serving system optimized for many-adapter environments.
Chameleon
introduces two new ideas: adapter caching and adapter-aware scheduling. 
First, 
Chameleon caches 
popular adapters in GPU memory, minimizing adapter loading times.
For caching, it uses  otherwise idle GPU memory, avoiding extra memory costs. 
Second,
Chameleon uses a non-preemptive multi-queue scheduler
% mechanism
to efficiently account for workload heterogeneity.
% heterogeneity in input/output size and adapter rank, efficiently. 
%managing workload heterogeneity.
In this way, Chameleon simultaneously prevents head of line blocking and starvation. 
Under high loads, 
Chameleon reduces the P99 and P50 TTFT latencies by 80.7\% and 48.1\%, respectively,
over a state-of-the-art baseline,
while improving the throughput by 1.5$\times$. 
\end{abstract}

%\thispagestyle{empty} % No page number

%\input{Source/cover-page}

%\clearpage



% === MAIN PAPER STARTS HERE ===
%\pagenumbering{arabic} % Start numbering pages
%\setcounter{page}{1}   % Start from page 1

%\pagenumbering{empty}
\pagenumbering{gobble}

\maketitle

\vspace{-3mm}
\section{Introduction}

%Generative 
Generative Large Language Models (LLMs)
have seen an exponential growth in recent years
% , triggering a long line of research~
\cite{orca,vllm,flashattention,spotserve,pets,dynamollm,splitwise,alizadeh2024llm,alisa}.
They have become integral to 
numerous technologies and 
applications~\cite{llmHealth,wiki:copilot,copilot,analytics,education}.
% , e.g. healthcare~\cite{llmHealth}, coding~\cite{copilot}, data analytics~\cite{analytics}, or education~\cite{education}.  
As their popularity increases, the number of online queries received by datacenter inference clusters   continuously grows~\cite{statChat}. These queries typically target a variety of downstream tasks, e.g., chat-bot conversation, coding, or text summarization.
These different tasks require different or special-purpose fine-tuned LLMs to achieve their highest accuracy.
Unfortunately, 
this requirement imposes a large hardware~\cite{splitwise} and energy~\cite{towardsGreen} tax on datacenters, as each of these models typically requires 
% vast amounts of
large memory  and, thus, many GPUs, to store  its many parameters. 

To alleviate this problem, adapter-based techniques  such as Low-Rank Adaptation (\emph{LoRA})~\cite{lora,lorapro}, have been explored.
These methods fine-tune   a small 
%(low-rank) 
subset of a base model's parameters for every task.
% , and have been originally proposed to speed up LLM training. 
Recent 
% inference
serving systems~\cite{slora,punica} leverage this technique.
They 
decouple the base model and the fine-tuned adapter parameters,
allowing different colocated LLMs to share   the  base model. 
This  
enables serving  potentially hundreds of LoRA fine-tuned LLMs at a
much lower memory cost.

% Despite the memory footprint reduction in multi-task settings, 
However, our characterization of this environment  shows that  
adapter-based LLM serving systems exhibit
two   challenges that substantially reduce  performance.
First, inference clusters have to orchestrate the adapters required by  incoming requests as they are being scheduled. 
% Punica~\cite{punica} and S-LoRA~\cite{slora} 
State-of-the-art systems~\cite{punica,slora}
keep the base model 
% always 
stored in GPU memory and the adapters in  host memory. 
% They consider storing adapters in the GPU memory too expensive as it can interfere with the memory allocations of incoming requests.
% , i.e. for the key-value caches of their intermediate state.
Then, they 
% propose 
fetch on-demand the adapters required by the running requests and discard them from the GPU memory as soon as the requests terminate.
%This synchronous loading penalizes performance, and, thus, 
% S-Lora~\cite{slora} and dLora~\cite{dlora-osdi} 
Some systems~\cite{slora,dlora-osdi}
further fetch in advance the adapters for the requests waiting in the system's queue to hide some of the loading overheads. 
However, 
our study reveals that  even such
asynchronous adapter fetching 
increases the time-to-first-token (TTFT) latency, especially when the system is heavily loaded, as it increases  contention in the CPU-GPU PCIe link. % bandwidth.
%Moreover, this data movement causes GPU stalls, resulting in low resource utilization.

Second, execution with adapters increases workload heterogeneity.
This is because decoupled computations between base model and adapters 
increase the execution time of  individual requests~\cite{dlora-osdi}, and such effect varies
across requests. Moreover, the use of adapters can increase
resource utilization and throughput, which results in the 
execution of  heterogeneous batches of requests for different tasks 
and  adapters~\cite{slora,dlora-osdi,punica,caraserve}.
With increased heterogeneity, tail latency is penalized: large requests that take long to execute end up
stalling  smaller requests within the same batch~\cite{caraserve}.

\begin{comment}
To reduce this bottleneck,
one can cluster the requests for the same adapter~\cite{dlora-osdi} or for adapters of the same rank~\cite{caraserve} within the same node, i.e., server in a cluster that stores a given model and adapter replicas, potentially causing load imbalance and frequent request migration~\cite{dlora-osdi}. 
While focusing on multi-node distributed scheduling,
most systems fail to account for the inevitable workload heterogeneity
within a replica, i.e., at \emph{server-level}.
\end{comment}
% Our study corroborates previous observations that LLM inference requests are prone to head-of-line blocking~\cite{uServe,fastserve}. 

\begin{comment}
OLD Second, %apart from the loading costs,
adapters introduce inference overheads and cause high workload heterogeneity.
Specifically, decoupled computations between  base model and adapter 
 % weights
increase the execution time of an individual query~\cite{dlora-osdi}. 
Moreover, the execution of a heterogeneous batch~\cite{slora,dlora-osdi,punica,caraserve}, 
% introduces optimized CUDA kernels that
% executes batches of heterogeneous requests, 
i.e., of requests for different tasks and LoRA adapters, 
% and thus
increases resource utilization and throughput, but penalizes tail latency:
% Specifically, 
requests for larger adapters take longer to execute and  stall the execution of smaller requests within the same batch~\cite{caraserve}.
\end{comment}

We analyze real-world production workloads~\cite{splitwise} and observe that requests follow a heavy-tailed distribution: most are completed in a  short time, while a small fraction experiences significantly longer execution durations.
While prior work has largely attributed this heterogeneity to differences in input~\cite{fastserve} and output~\cite{uServe} request sizes, 
% our findings uncover additional critical factors,
our study is the first one to shed light on how the variability in adapter rank (i.e., size)~\cite{slora} and popularity~\cite{serverlessllm,dynamollm,dlora-osdi} affect the requests at the tail, underlying the necessity to take the adapter size into account.

Unfortunately, simply prioritizing short requests is insufficient to address the issue of tail latency. 
For instance, the speculative Shortest-Job-First (SJF) scheduler~\cite{uServe}, %employed by $\mu$Serve~\cite{uServe}, 
along with its aging mechanism to mitigate starvation, inadvertently increases the tail latency of longer requests---potentially causing them to miss their Service Level Objectives (SLOs).
Instead, our findings emphasize the need for a more nuanced scheduling strategy: one that addresses adapter-level heterogeneity, offers expedited processing for short requests, and ensures that longer requests still meet their SLOs.

\begin{comment}
JOVAN: PREVIOUS VERSION
% To that end, 
Our study corroborates past findings that LLM inference requests suffer from head-of-line blocking~\cite{uServe, fastserve}. Specifically, we show that requests from real production environments~\cite{splitwise} follow the heavy-tail pattern, i.e. the majority of the requests have very short execution time, while a few requests have very long execution time.
% but there are some long requests as well. 
Prior art considers only the effect of the varying input~\cite{fastserve} and output~\cite{uServe} sizes for this heterogeneity. 
Our study is the first to shed light on how the variability in adapters rank~\cite{slora} and popularity~\cite{serverlessllm,dynamollm,dlora-osdi} affect the requests at the tail, underlying the necessity to take them into account. 
We also show that aggressively prioritizing short requests does not solve the problem, i.e. $\mu$Serve's~\cite{uServe} speculative Shortest-Job-First (SJF) scheduler and the aging mechanism it employs to handle starvation increase the tail latency of longer requests. \emph{A scheduling mechanism that takes into account adapter heterogeneity and provides a fast lane for short requests while guaranteeing that longer requests execute under SLOs is desired.}
\end{comment}

We use these insights to design \emph{\system{}}, an  LLM 
inference serving system optimized for 
many-adapter
environments.
Tasks share their base LLM, which uses a large fraction of the GPU memory, while
each task uses its own specific adapter.
%, which occupies a small fraction of GPU memory. 
\system{} attains high efficiency through two new ideas. 

%First, \system{} provides a  transparent, adaptive and interference-free cache for adapters. 

First, Chameleon provides a {\em transparent, adaptive, and interference-free cache for adapters}.
Contrary to common wisdom~\cite{slora,punica}, we observe that, even during high load,
there is enough idle GPU memory  to implement a cache for adapters that are likely to be reused in the future. 
However, as available memory fluctuates, the cache must be dynamically sized and carefully managed to avoid interfering with the key-value cache, while employing a cost-aware eviction policy suited for workload heterogeneity.

Second, \system{} employs a \emph{non-preemptive, adapter-aware multi-level queue (MLQ) scheduler} to minimize head-of-line blocking and ensure SLO compliance for all request types.
Requests are classified into different queues based on their predicted sizes  and, in each scheduling cycle, a subset from each queue is selected to form a batch.
This  enables a faster lane for smaller requests while also  eliminating starvation  across all request sizes.

\begin{comment}
Prior art considers feedback queues (MLFQ) to schedule LLM inference requests, and relies on preemption to deal with their non-deterministic execution times~\cite{fastserve}. This requires  the orchestration of the intermediate states of preempted requests, introducing non-negligible complexities. 
\system{} instead speculates a \emph{weighted request size (WRS)} and uses it to assign 
priorities, i.e. admit requests to specific queues. 
WRS takes into account the number of input tokens,
an estimated output length~\cite{uServe} and 
the adapter rank for every request.
Unlike preemptive solutions, 
\system{} admits requests from all queues to every batch but partitions resources in proportion to the queues priority.
Specifically, 
\system{} assigns a different resource quota to each queue~\cite{qzilla}, i.e. a maximum number of tokens it can admit to a batch, 
and gradually decreases it for the queues used by larger requests. 
This \emph{enables a faster lane for smaller requests} but also \emph{eliminates starvation}.
To guarantee SLOs
for all requests in every queue and  
maximize throughput, \system{} dynamically adjusts the number of queues and their resource quotas at runtime.
\end{comment}

We implement \system{} on top of the open-source S-LoRA~\cite{slora} LLM serving platform. 
\system{} does not require any hardware or
operating system  support, or changes to CUDA kernels.
We evaluate \system{} with open-source LLMs using real-world production traces~\cite{splitwise}
and show that \system{} is very effective.
Compared to a state-of-the-art
baseline~\cite{slora}, \system{} 
reduces the P99 and P50 time-to-first-token (TTFT) latencies by 
80.7\% and  48.1\%, respectively, 
while improving the throughput by 
1.5$\times$.

This paper makes the following contributions:

\begin{itemize}[topsep=0pt, leftmargin=*]
\item A characterization of state-of-the-art LLM inference serving
systems in environments with many LoRA adapters.
\item The \system{} LLM inference serving platform, which  introduces the first cache design for LoRA adapters, and a novel adapter-aware multi-queue scheduler that eliminates head-of-line blocking while preventing starvation.
\item An implementation and evaluation of \system{}.
\end{itemize}
\vspace{1pt}
\vspace{-3mm}
\section{Background}
\label{sec:background}

%\nikoleta{RESUMED FOR CAMERA}

%\nikoleta{BEGIN}
%CAMERA MICRO 25  

\myparagraph{LLM inference}
Generative LLMs~\cite{llama2, llama3, mixtral2,radford2019gpt} 

process the entire input at once (\emph{prefill phase}) and then 
generate output tokens 
%auto-regressively, i.e.
one by one (\emph{decode phase}). 
In prefill, all input tokens are processed in parallel. This phase is compute-bound and its performance depends 
on the input size, which is  known in advance.
In decode, the output tokens are generated sequentially in iterations. 
Each iteration generates a token based on the input prompt and all previously generated tokens, typically cached on the GPU memory in
\emph{key-value (KV) caches}.

The decode phase is memory-intensive and 
its performance depends on the output size, i.e. the number of decode iterations, which is  determined on the fly and  is unknown at the time a request is admitted to execute.

\myparagraph{LLM inference serving systems} LLM   serving systems batch  requests for the same model to maximize hardware utilization.
Since different requests generate different numbers of output tokens, the execution time of different requests in the same batch varies. 
To prevent long requests from blocking smaller ones, 
systems dynamically update batches.
Specifically, state-of-the-art systems perform
continuous 
batching~\cite{orca,sarathi}: they remove completed requests from a batch and potentially add new ready-to-run requests on every decode iteration, an approach called iteration-level scheduling.

\myparagraph{LLM Low-Rank Adaptation (LoRA)}
One way to reduce LLM training overhead is to pre-train LLMs and then fine-tune them for specific
tasks. One method to do so is Low-Rank Adaptation (LoRA)~\cite{lora,adapterhub}, where the 
layers of a base model are updated with low-rank matrices to fine tune them. These
matrices are called LoRA adapters and their size (i.e., their {\em rank}) determines the accuracy of
the resulting 
computation. Specifically, higher ranks potentially translate to better tuning and higher accuracy.
Since different tasks have different accuracy requirements, they are likely to employ adapters of different ranks over the same base LLM~\cite{lorapro,slora}. 

The straightforward way to 
apply adapters is to merge them with the base model and create a full-size standalone specialized LLM instance~\cite{peft} for each task.  
However, recent works for LLM inference serving~\cite{slora,punica,dlora-osdi,caraserve} 
allow tasks to share the  base model and allocate specific per-task  adapters.
% instead of having standalone per-task merged instance(s). 
Typically, an adapter is significantly smaller than the base model. Hence,  
this method significantly reduces the memory requirements of  systems that serve  diverse tasks~\cite{pets}. 
Moreover, such method also enables batching  requests of different tasks, with a common base and different adapter combinations, further improving the throughput.

\begin{figure}[t]
% \vspace{-2mm} 
\centering \includegraphics[width=0.8\columnwidth]{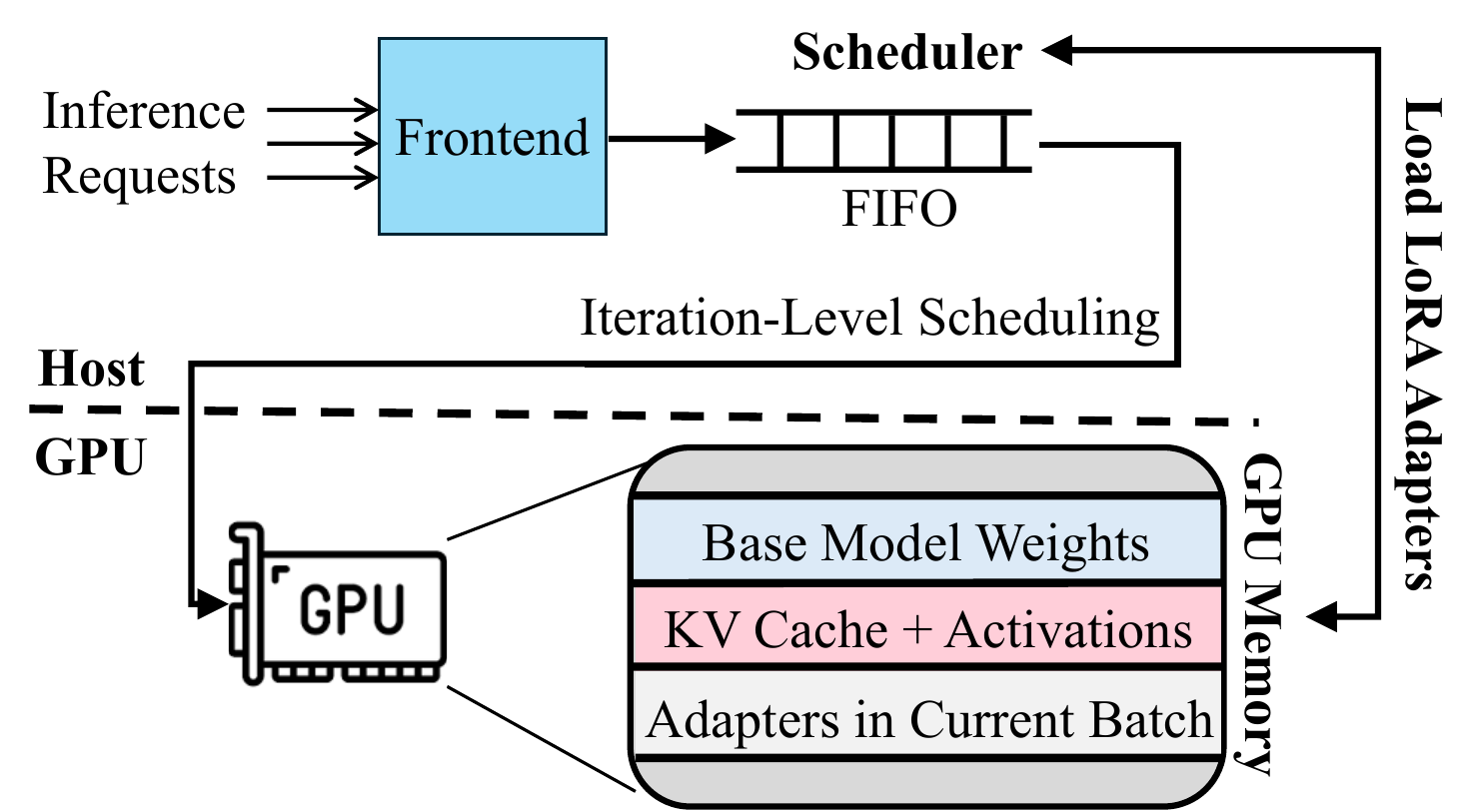}
\vspace{-3mm} 
\caption{Conventional LoRA online serving system.} 
\label{fig:backgroundlora}
\vspace{-3mm} 
\end{figure}

%Figure~\ref{fig:backgroundlora} shows the organization of conventional systems for LLM inference serving in the presence of LoRA adapters~\cite{slora,dlora-osdi,punica,caraserve}.
Figure~\ref{fig:backgroundlora} shows the organization of such   systems~\cite{slora,dlora-osdi,punica,caraserve}.
On initialization, the base LLM model is %loaded into the CPU memory and 
transferred to the GPU memory from the host.
A scheduler on the host manages the incoming requests, updating the batch to be executed on every iteration.
% (iteration-level scheduling) .
%On every LLM iteration,
%the scheduler creates an updated batch of requests and sends it for execution to the inference engine on the GPU.
Before it sends the batch to the inference engine on the GPU, the 
scheduler also loads any missing adapters required by the requests in the batch. 
%transfers to the GPU memory
%all the adapters needed by requests in the batch that are currently not present.
%in the GPU memory.
Once there are no 
%running 
requests that use a given adapter,
the adapter is discarded from the GPU memory to make room for new incoming requests~\cite{slora,punica}.

%\vspace{-1pt}
\section{Opportunities in Many-Adapter Settings}
%\section{Opportunities for Efficient LLM Serving in Many-Adapter Environments}
\label{sec:char}

In this section, we examine the new challenges that appear in  many-adapter environments  and why they are not efficiently handled by conventional  LLM serving systems.
We characterize the open-source Llama-7B model~\cite{llama2} on an NVIDIA A40 server with 48GB of GPU memory~\cite{a40}.
We use the S-LoRA serving platform~\cite{slora}, a state-of-the-art inference system for multi-adapter scenarios.

\subsection{Adapters Increase Workload Heterogeneity}
\label{sec:hetero} 
\begin{figure}[t]
\vspace{0mm}
\centering
\includegraphics[width=0.9\columnwidth]{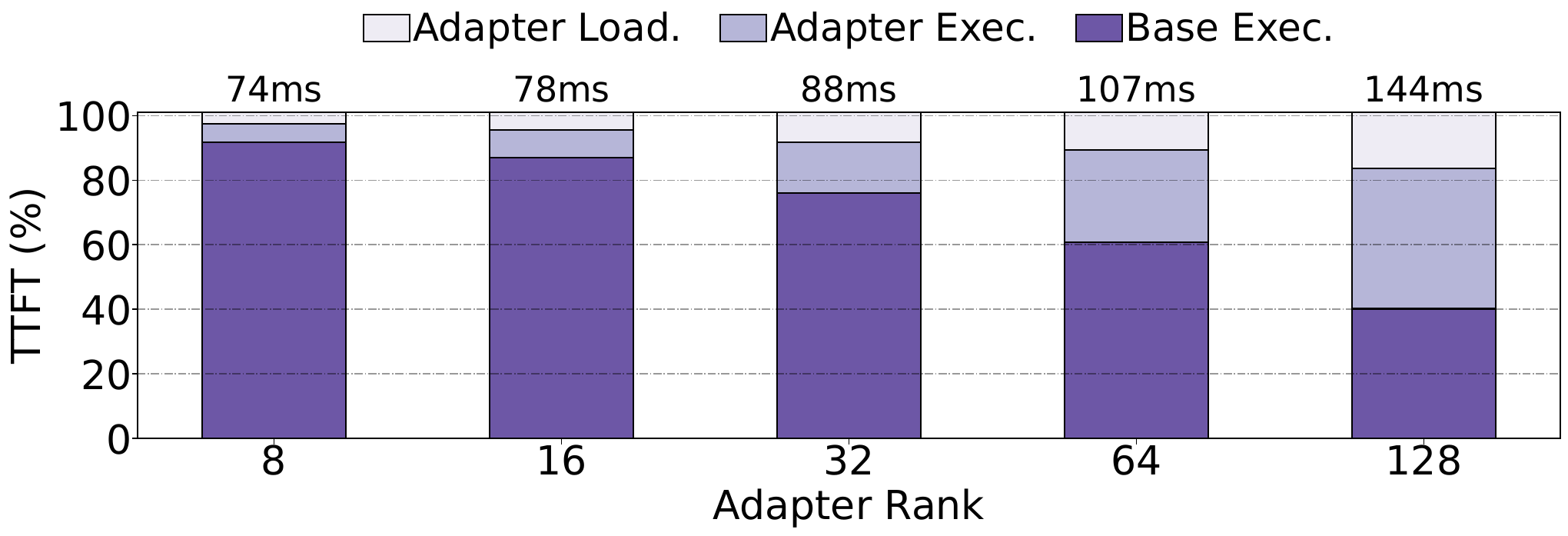}
\vspace{-4mm}
\caption{TTFT latency with different adapter ranks broken down into base and adapter execution, and adapter loading.}
\label{fig:adapter-load}
\vspace{-3mm}
\end{figure}

The LoRA adapters employed by different tasks  are expected to vary in size (\emph{rank}), as tasks require different levels of accuracy~\cite{lora,lorapro,slora,dlora-osdi}. 
Figure~\ref{fig:adapter-load} shows how this  rank heterogeneity  affects the TTFT of a single inference request, with medium input and output size~\cite{dynamollm}. We run the request  
over a base Llama-7B model combined with a specific adapter on an unloaded system, and increase the adapter rank from 8 to 128~\cite{dlora-osdi,slora}. 
We  break down the total execution time into time spent: (1) executing the  base model, (2) 
executing the adapter, and (3) loading the adapter's weights from host to the GPU memory. 
The numbers on top of the bars are the  TTFT values.

We observe that, as the rank size increases, the relative weight of the 
adapter overheads also increases.
For example, for rank 128, $\sim$60\% of the total TTFT latency is spent on adapter loading and computation.

For these experiments, we use the Multi-size Batched Gather Matrix-Matrix Multiplication (MBGMM) kernel from the state-of-the-art baseline system S-LoRA~\cite{slora}. LoRA adapters induce two matrix multiplications
%, Ax and BAx, 
on top of the base model multiplication, 
%Wx, 
and a matrix addition for results aggregation per LLM inference layer. This leads to the high computational overhead of adapter execution observed in Figure~\ref{fig:adapter-load}. Recent work (Figure 5 in~\cite{dlora-osdi}) corroborates our findings that these steps are expensive even for small-rank adapters.

\begin{figure}[t]
% \vspace{-2mm}
\centering
\includegraphics[width=0.9\columnwidth]{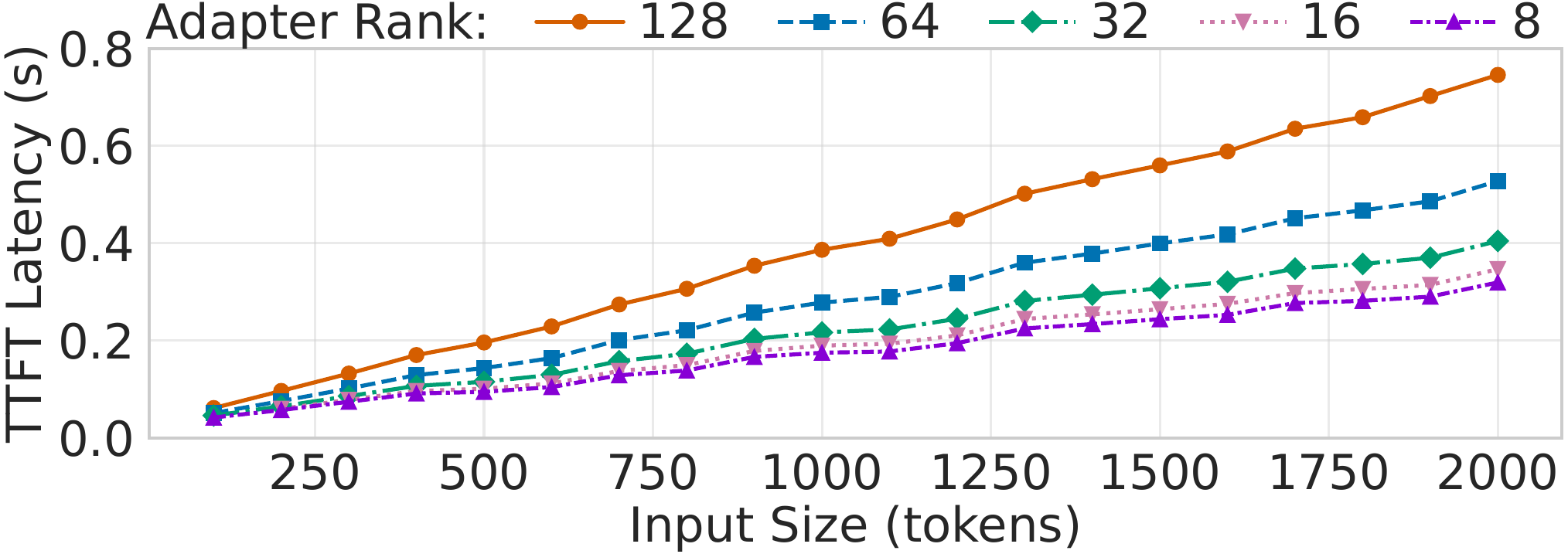}
\vspace{-3mm}
\caption{TTFT latency for different adapter ranks.}
% while varying the request's input size.}
\label{fig:adapter-ttft}
\vspace{-5mm}
\end{figure}

We further examine the effect of the adapter rank while considering other sources of inference heterogeneity. 
Prior work observed that large inputs lead to longer prefill phases and large outputs to much longer decode phases~\cite{dynamollm}. 
Also, using large batches of requests increases throughput but at the cost of longer decode iterations.
Figure~\ref{fig:adapter-ttft}
shows the TTFT latency
for different adapter ranks as we vary the 
input size of a request (i.e., the number of input tokens), while keeping the output size fixed.
For this experiment, we keep the adapter weights in GPU memory and isolate prefill performance by
excluding   adapter loading.
For all input sizes, TTFT  varies significantly across  adapter ranks. Moreover, the impact 
of  the rank is more pronounced as the input size increases. 
Similarly, it can be shown that, for large batch sizes, different adapter ranks lead to diverse decode latencies for requests with similar input/output sizes. 
Overall, we find adapter rank to be an extra, equally important source of heterogeneity, next to input, output, and batch size.

Apart from   different ranks, adapters  have skewed popularity as well, following the skewed popularity of different tasks.
LLM inference is a user-facing service
where some tasks receive a  larger amount of requests than others, and these requests typically arrive in bursts~\cite{dlora-osdi,serverlessllm,dynamollm}. 
Next, we will show how this heterogeneity affects various system design decisions, such as which adapter to keep in GPU memory or how to schedule inference requests for different adapters.

\vspace{1pt}
\noindent \textbf{Insight \#1:} Adapters are an additional source of heterogeneity in LLM inference that must be managed dynamically.

\subsection{Adapters are Expensive to Load}
\label{sec:load}
When an LLM inference request using a specific adapter arrives at an online serving system, the adapter 
%weights 
must be loaded into the GPU memory for the request to be processed. 
Thus, loading the adapter weights lies on the critical path of inference execution.  
Figure~\ref{fig:adapter-load} shows that 
loading takes  17.5\% of the total TTFT latency when a 128-rank adapter is used in an unloaded system.

\begin{figure}[h]
\centering
\includegraphics[width=0.9\columnwidth]{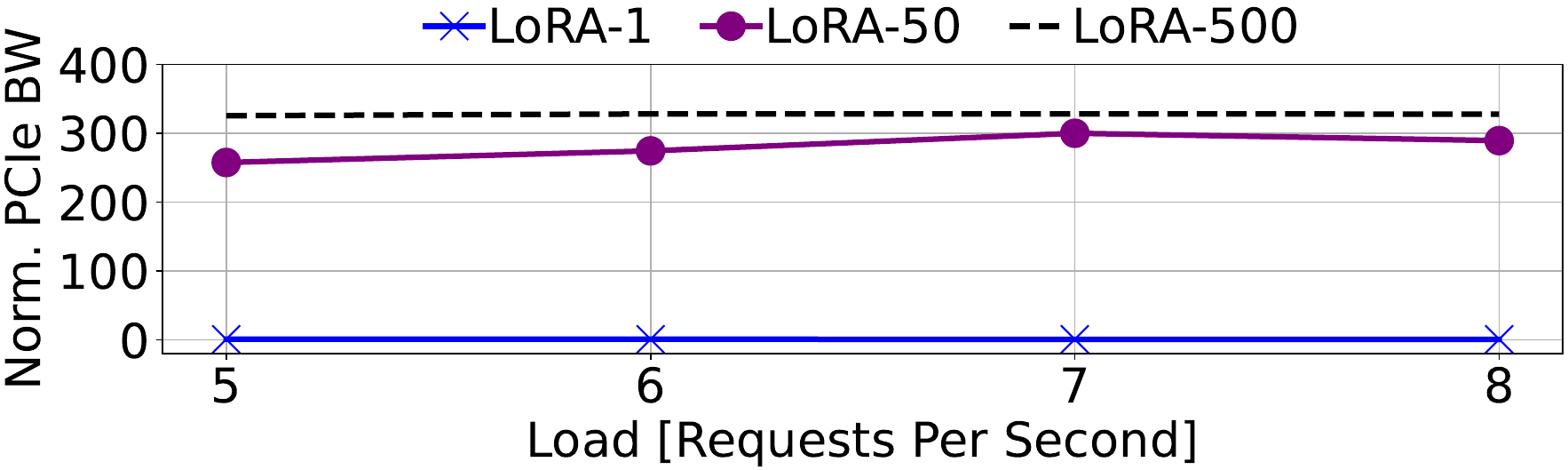}
\vspace{-4mm}
\caption{PCIe bandwidth usage under different loads for environments with: 1 adapter (\emph{LoRA-1}), 50 different adapters (\emph{LoRA-50}), and 500 different adapters (\emph{LoRA-500}).
% \todo{Include also latency results -- use secondary axis}
}
\label{fig:adapter-pcie-new}
\vspace{-3mm}
\end{figure}

This overhead  becomes more pronounced as the requests use a larger number of different adapters.
One reason is the contention on the PCIe link between the host and the  GPU as   
the adapters are brought to the GPU memory. In our next experiment, we use rank 32 adapters and
consider three scenarios: in \emph{LoRA-1}, all the 
requests use the same adapter; in \emph{LoRA-50} and \emph{LoRA-500}, a request uses one of 
50 or 500 different adapters with a uniform distribution. 
Figure~\ref{fig:adapter-pcie-new} shows the normalized PCIe bandwidth consumption 
for the three scenarios and different requests per second (RPS).
We normalize the bandwidth consumption to \emph{LoRA-1} with  5 RPS.

We see that, as we go from \emph{LoRA-1} to \emph{LoRA-50} and \emph{LoRA-500}, the
bandwith consumption increases. With \emph{LoRA-500}, the PCIe bus is saturated.
We measure that at, 8 RPS, this bandwidth contention causes the  P99 TTFT latency 
of the requests in \emph{LoRA-50} and \emph{LoRA-500} to be  1.69x and 2.60x higher, respectively, than
\emph{LoRA-1}. For higher RPS loads, these P99 TTFT latency gaps increase rapidly, but they are also
affected by other bottlenecks.

%beyond loading the adapters.

\begin{comment}
    These overheads are more pronounced when the system is loaded, i.e., the TTFT latency of LLM inference requests deteriorates as the number of  adapters used by the requests increases.
%The main 
One reason is the contention on the PCIe link between the host and the  GPU. Figure~\ref{fig:adapter-pcie-new} shows the normalized PCIe bandwidth consumption at 
different loads 
%\hl{and its effect to P99 TTFT latency}
when we increase the number of rank-32 adapters used by the requests.
Specifically, 
we evaluate systems with a single adapter used by all requests (\emph{LoRA-1}), and with 50 or 500 different adapters uniformly distributed across requests (\emph{LoRA-50} and \emph{LoRA-500}).
We normalize the bandwidth consumption to that of the \emph{LoRA-1} environment at 5 
requests per second (RPS).

We observe that, as the number of adapters increases,
the PCIe bandwidth usage grows significantly.
\hl{We measure that for LoRA-50 and up to 10 RPS, this BW contention increases P99 TTFT latency by 11.8$\times$.}
Interestingly, at higher loads, \emph{LoRA-50} becomes compute-bound, as most of the 50 adapters remain loaded in the GPU memory by the running requests and thus 
%due to the queuing effects, 
the PCIe bandwidth consumption drops.
In contrast, \emph{LoRA-500} is constantly bound by the PCIe bandwidth: 
\emph{LoRA-500} has 328$\times$ higher PCIe bandwidth consumption than \emph{LoRA-1}
and 48.1$\times$ higher TTFT latency at 10RPS. 
%\XA{After all I am not that sure how these can be merged to one figure.}

\end{comment}

We now evaluate the impact of adapter loading overhead  when serving the larger Llama-70B base model using tensor parallelism (TP) across 2, 4, and 8 A100 GPUs—since the model no longer fits on a single GPU. We observe that the cost of loading adapters increases due to two main factors. First, larger base models result in proportionally larger adapter weight matrices, for the same rank configuration, increasing their loading time.
%the size of each adapter and thus the loading time. 
For example, a rank 32 adapter for Llama-7B is 
64 MB, while its size grows to 256 MB for Llama-70B.  
Rank 128 adapter size grows to the order of GBs.  
%For example, it can be shown
%Figure~\ref{fig:large-model} 
%Thus when running LLaMA-7B on TP2 with adapter rank 8, adapter loading accounts for 21\% of the TTFT latency, while LLaMA-70B on TP2 with the same rank consumes 30\%. 
Second, using more GPUs introduces additional overheads: adapter weights must now be partitioned across tensor-parallel ranks, transferred separately to each GPU’s memory, and synchronized to ensure consistent execution. These overheads 
%in adapter sharding and communication 
exacerbate the latency on the critical path of inference. %For instance, it can be shown 

\begin{figure}[t]
\vspace{0mm}
\centering
\includegraphics[width=0.9\columnwidth]{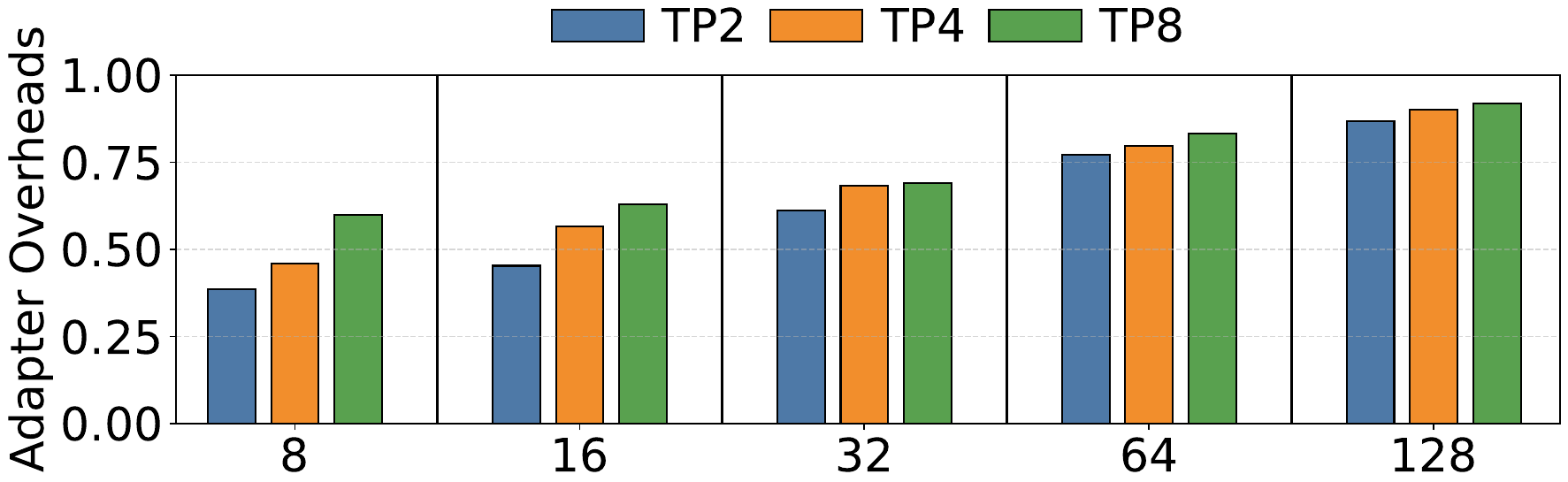}
\vspace{-4mm}
\caption{Overhead of loading the adapters of different
ranks as a fraction of the total TTFT latency for Llama-70B running on 2,
4, or 8 A100 GPUs using tensor parallelism.}
\label{fig:large-model}
\vspace{-2mm}
\end{figure}

Figure~\ref{fig:large-model} considers different TP degrees and adapter ranks, and shows
the fraction of the TTFT latency that is taken by adapter loading. We observe that this fraction  increases
with the TP degree  and adapter rank. For example, loading accounts for 68\% of the TTFT latency
for rank 32 and TP4.

%shows how these factors 
%make adapter loading costlier for large models and under Tensor Parallelism. For example, load latency accounts for 43\% of the TTFT latency even for rank size 8 when running Llama-70B on TP 4. 

%Figure~\ref{fig:large-model}  shows how these factors 
%make adapter loading costlier for large models and under Tensor Parallelism. For example, load latency accounts for 43\% of the TTFT latency even for rank size 8 when running Llama-70B on TP 4.  
%that these factors make 
%adapter loading 
%that adapter loading accounts for 43\% of the TTFT latency when running LLaMA-70B on TP8, compared to 30\% on TP2 with the same adapter rank.

\begin{comment}
\begin{figure}[t]
\centering
\includegraphics[width=\columnwidth]{Source/Source/figures/characterization/lat_pcie_log2_crop.pdf}
\vspace{-6mm}
\caption{\hl{P99 TTFT under different loads for environments with: 1 adapter (\emph{LoRA-1}), 50 different adapters (\emph{LoRA-50}), and 500 different adapters (\emph{LoRA-500}).}
}
\label{fig:adapter-pcie-lat}
\vspace{-3mm}
\end{figure}
\end{comment}

\begin{figure}[t]
\vspace{0mm}
\centering
\includegraphics[width=0.9\columnwidth]{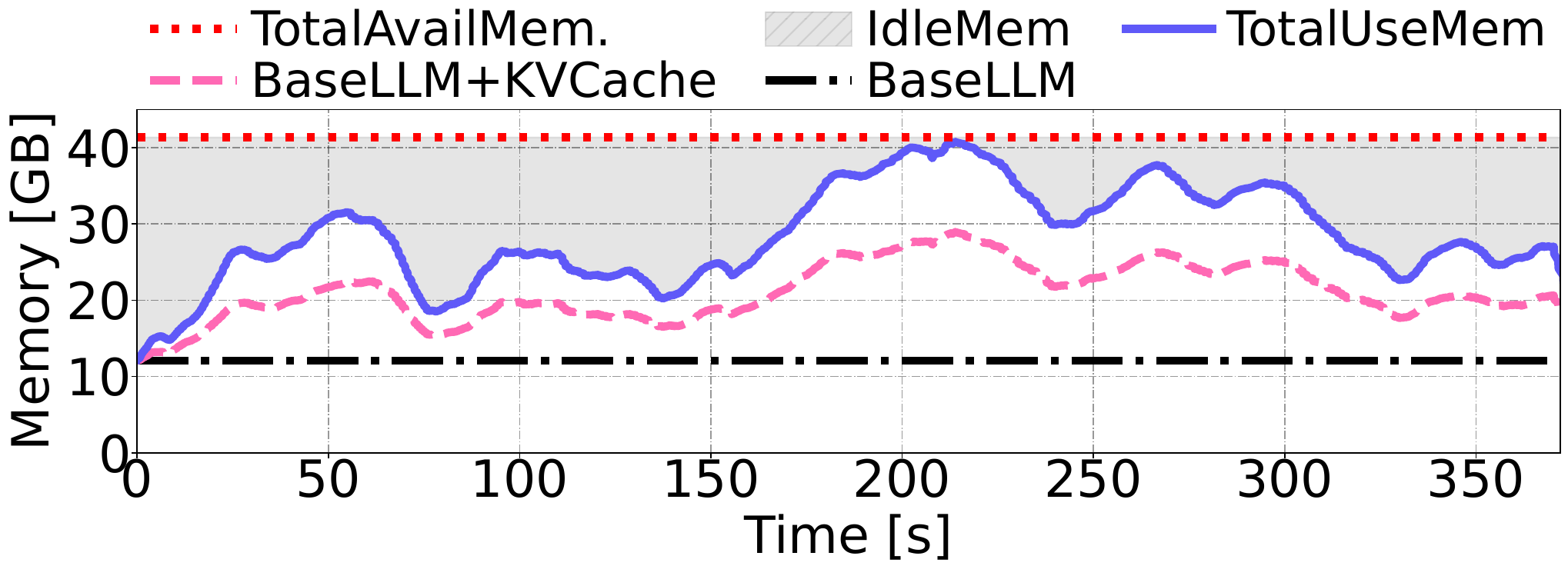}
\vspace{-4mm}
\caption{Memory usage over time for different parts of the workload: base LLM model, KV cache, and adapters.}
\label{fig:mem-time}
\vspace{-5mm}
\end{figure}

%, i.e. base model, KV cache and requests input and output, 

An intuitive way to reduce these overheads is to leverage idle GPU memory to cache %recently used 
adapters.
However, LLM inference
has substantial load fluctuations~\cite{dynamollm}.
Figure~\ref{fig:mem-time} shows the GPU memory usage over time when we run the Llama-7B model using production traces of requests from Azure~\cite{splitwise}. Because our testbed has modest memory,
we have scaled down the input and output lengths in
 these large-scale system traces 
 %linearly, 
 using a constant factor that results in the 
peak memory consumption of the scaled-down trace to be equal to the memory capacity of our testbed (\S~\ref{sec:methodology}).

The figure shows the memory consumed by the base LLM ({\em BaseLLM}), 
base LLM plus KV cache ({\em BaseLLM+KVCache}), and
base LLM plus KV cache plus adapter ({\em TotalUseMem}).
We see that, most of the time, there is abundant idle memory that can cache adapter weights.
However, idle memory drastically drops during load spikes.
Hence, the system needs to carefully and dynamically resize the cache resources based on the incoming load. %employing a  cost-aware eviction policy  taking into account per-adapter reloading cost.

These findings challenge the common design decision to discard the adapters from GPU memory if none of the currently running requests use them~\cite{slora,dlora-osdi}.
We find that keeping them in GPU memory can significantly improve performance, especially in high load scenarios,
and 
that there is a sizable amount of idle GPU memory that can be used for this.

\vspace{3pt}
\noindent \textbf{Insight \#2:}
Frequent loading of adapters from host to GPU memory creates bandwidth contention, degrading   system performance. Idle GPU memory can be repurposed to cache adapters and mitigate some of these overheads. However, dynamic resizing of the cache is essential, as the amount of idle memory fluctuates heavily.
%based on request traffic. 

\subsection {Adapters Affect Requests at the Tail}
\label{sec:tail}

In Section~\ref{sec:hetero}, we 
observed that there is a high degree of heterogeneity in the performance of LLM inference requests, based on their input, output, and adapter size.
Now, we analyze how this heterogeneity impacts the effectiveness of scheduling decisions.

\begin{figure}[t]
\centering
\includegraphics[width=\columnwidth]{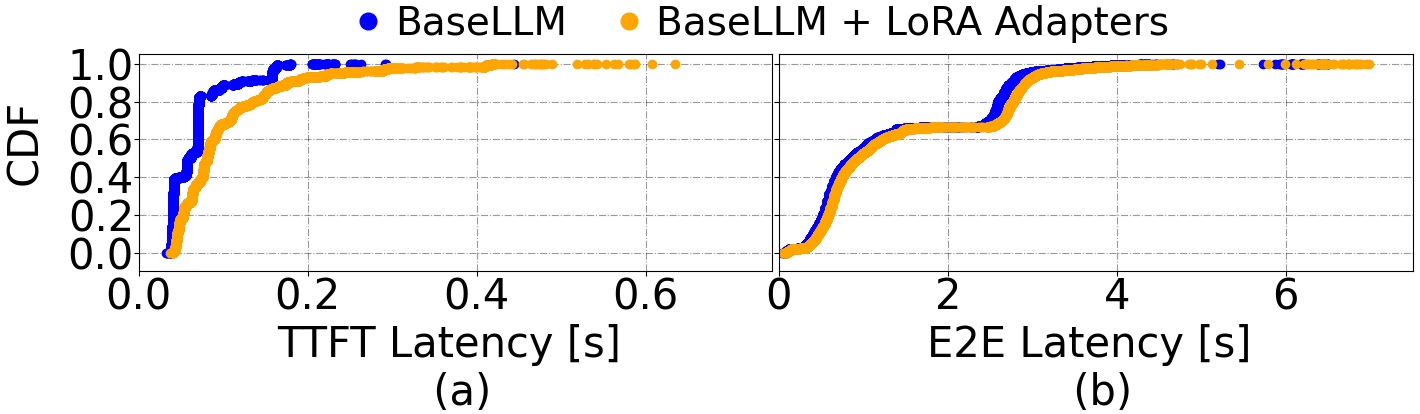}
\vspace{-8mm}
\caption{CDF of (a) TTFT and (b) E2E latency of requests for a real LLM trace \cite{splitwise}. Requests are executed one by one.}
\label{fig:adapter-trace}
\vspace{-6mm}
\end{figure}

We take the open-source production traces of LLM inference requests for a conversation service~\cite{splitwise} and execute one request at a time. 
We run with only a base LLM, and with a base LLM augmented with LoRA adapters.
Similar to ~\cite{slora}, we consider a pool of 100 different adapters with rank sizes uniformly distributed among 8, 16, 32, 64, and 128, and %~\cite{slora}. 
associate every request in the trace with one of these adapters, following %the 
a uniform %power-law 
distribution 
for   rank popularity
and a power-law distribution for  adapter popularity.
%We follow a power-law distribution for the popularity of the adapters.
%~\cite{slora}.
%Adapters with smaller ranks are the ones more frequently requested and we distribute uniformly the requests among the adapters with the same rank.
%For adapters of the same rank we distribute requests uniformly.
Figure~\ref{fig:adapter-trace} shows the CDF of (a) TTFT and (b) end-to-end latency of all requests.
For this experiment, the latency includes both the prefill phase and the time it takes to load the adapter.
This figure shows that the execution time of requests follows a heavy-tail pattern: the majority of requests have  short execution times, but there are a few very long requests.
Moreover, adding LoRA adapters significantly affects requests at the tail.

%, in a real production environment, requests remain highly heterogeneous and their execution time follows the heavy-tail pattern: the majority of requests have very short execution times, but there are a few very long requests.
%Importantly, adding LoRA adapters significantly affects requests at the tail. 
%\hl{In section} ~\ref{sec:eval} 
%\hl{we report performance results
%for a varying total number of adapters and for power-law rank popularity showing that these findings are persistent.}

Heterogeneity in execution times typically requires special scheduling considerations.
LLM engines schedule   requests at iteration-level~\cite{orca}  where, at each iteration, the scheduler decides which requests will execute in a batch.
The majority of conventional systems use a FIFO approach due to its simplicity~\cite{slora,vllm}.
However, FIFO is inefficient for heterogeneous requests, as it
introduces head-of-line (HoL) blocking, leading to increased tail latency. 

For this reason, researchers have proposed to schedule the requests in a Shortest-Job-First (SJF) manner. Specifically,   existing systems~\cite{uServe} predict the request output lengths and prioritize the requests with the shortest predicted outputs. 
However, continuously prioritizing short requests leads to the starvation of long requests, again, negatively impacting the overall tail latency.
Moreover, using the output length as the only scheduling knob is insufficient,
as inputs and adapters also impact the total latency
(Figure~\ref{fig:adapter-ttft}).

\begin{figure}[t]
\centering
\includegraphics[width=\columnwidth]{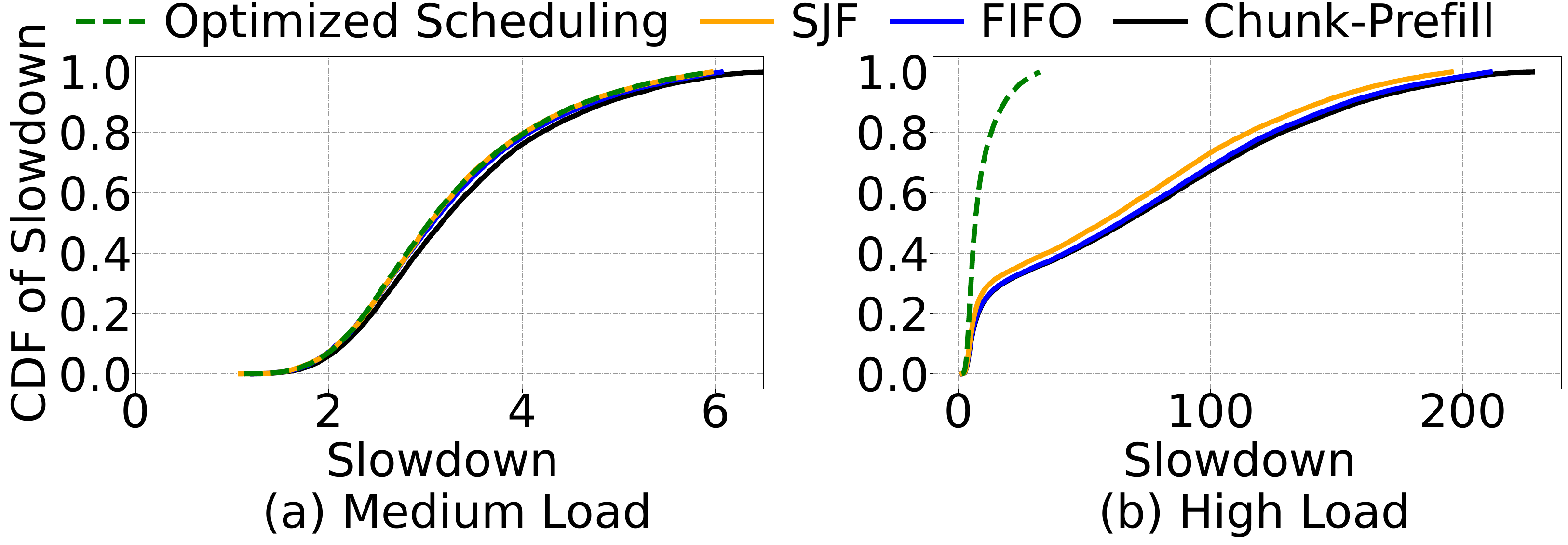}
\vspace{-7mm}
\caption{CDF of the per-request slowdown for different scheduling policies under (a) medium and (b) high load.}
\label{fig:slowdown}
\vspace{-7mm}
\end{figure}

To show the inefficiencies of these two  scheduling policies, we execute the production trace~\cite{splitwise} using the Llama-7B model.
We record the slowdown of each request: how 
many times higher is  the request's response time now relative to the response time in an 
isolated environment where the request executes alone.
Figure~\ref{fig:slowdown} shows the CDF of slowdown per request with different scheduling policies:
FIFO with regular iteration-level scheduling (i.e., continuous batching)~\cite{orca} ({\em FIFO});
FIFO with the more advanced chunked-prefill iteration-level scheduling~\cite{sarathi} ({\em Chunk-Prefill});
SJF; and the optimized scheduling policy that we will introduce in Section~\ref{sec:design} 
({\em Optimized Scheduling}). The last two schemes use iteration-level scheduling.

%FIFO and SJF scheduling policies, along with the optimized scheduling policy that we will introduce in Section~\ref{sec:design}.  For the FIFO policy we show results with regular iteration-level scheduling (i.e.,continuous batching)~\cite{orca} (FIFO) and with more advanced chunked-prefill iteration-level scheduling~\cite{sarathi} (Chunk-Prefill). The other schemes (Opt. Sched and SJF) use  iteration-level scheduling.

% While the scheduling policy does not play a significant role in medium loads,

Under high loads, conventional policies create high slowdowns for the requests at the tail. 
In FIFO, short requests are blocked by long requests. 
Using a classification of requests into short, medium, and long that
we describe in Section~\ref{sec:design}, we measure that 
a short request 
spends on average 28.6\% of its time waiting to be scheduled, compared to %16\% and 
12\% for a large request.
%for the rest 
%of the requests. 
As chunked-prefill is designed to prioritize decode iterations, it slightly slows down  prefill iterations, increasing   TTFT latencies. Chunked-prefill does not 
%fully 
solve the HoL blocking problem because it still adheres to a per-request ordering within each pipeline stage. Thus, short requests can remain blocked behind long prefill or decode chunks in their respective queues, especially when resources (e.g., tokens or compute slots) are saturated. Hence, chunked-prefill 
does not reduce tail latency under high-load scenarios.

%does not dynamically reorder requests across stages based on criticality or expected duration, limiting its ability to reduce tail latency under bursty or skewed load conditions.}
For SJF, long requests are penalized, as they are starved by the prioritization of short ones. We measure that a long request
spends 5.15 s waiting to be scheduled compared to 1.5 s 
for a small request.
%and 3.6 s
%for the rest of the requests.
While long requests are relatively infrequent, their queuing latency has
a significant   impact.
 \begin{comment}
\hl{For FIFO, short requests are blocked by long requests, making queuing delays account for} \todo{X\%} \hl{of the average request runtime. For SJF, long requests are penalized, as they are starved by the prioritization of shorter requests, and while they are less frequent they still make average queuing delays cost} \todo{X\%} \hl{of the average request runtime.}
For FIFO, these are the short requests blocked by the long requests, while for SJF, these are the long requests starved due to the prioritization of short requests.
\end{comment}

%An optimized scheduling policy can efficiently handle this inter-request heterogeneity and significantly improve the performance.

\vspace{1pt}
\noindent \textbf{Insight \#3:}
Conventional scheduling policies  such as FIFO and SJF  are ineffective for highly heterogeneous LLM inference requests. There is a need for a scheduling policy that can efficiently manage request heterogeneity 
while, at the same time, taking into account all knobs that affect the execution time.

\vspace{-2mm}

\section{\system{} Design}
\label{sec:design}

\subsection{Overview}
\label{sub_over}

Based on all the previous insights, % from our characterization,
we design \emph{Chameleon},
an LLM inference serving
system optimized for many-adapter environments.
Chameleon is designed to address the unique challenges posed by 
%the heterogeneity found in 
these environments:
(1) the overheads and side-effects of loading the adapters' weights and (2) the increased 
request tail latency  due to inefficient request scheduling. 

To address the first issue, Chameleon uses underutilized GPU memory to implement a software-managed adapter cache. This  \emph{Chameleon Adapter Cache}  
stores  adapter  weights in GPU memory, removing adapter loading off the
%the inference request's 
critical path, and reducing PCIe bandwidth usage. To address the second issue, Chameleon uses a multi-queue non-preemptive scheduler that provides an express lane for short requests, while ensuring that no request starves. This \emph{Chameleon Scheduler} works synergistically with the Adapter Cache,
maximizing system  throughput.

\begin{figure}[t]
% \vspace{-2mm} 
\centering \includegraphics[width=0.8\columnwidth]
{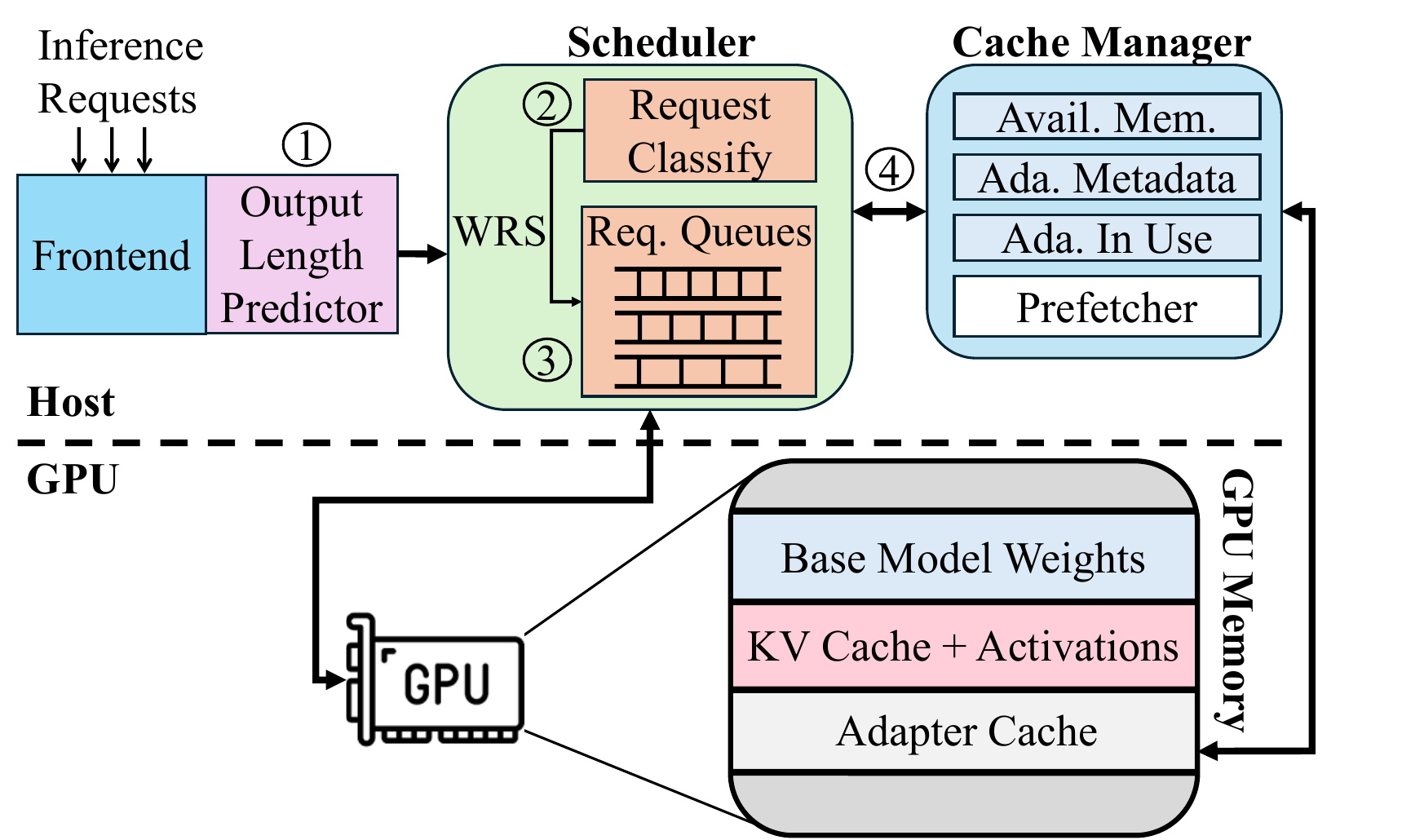}
%{Source/Chameleon_arch_jovan_crop.pdf} 
\vspace{-4mm} 
\caption{Chameleon architecture.} \label{fig:overview} 
\vspace{-6mm} 
\end{figure}

%\vspace{1pt}
%\noindent \textbf{Architecture Overview.}
Figure \ref{fig:overview} overviews the Chameleon architecture. 
To handle workload heterogeneity, Chameleon classifies all incoming requests based on their total size.
Every request first goes through an output length predictor that estimates the number of output tokens \circleint{1}.
We use an existing, open-source, predictor based on a BERT proxy model~\cite{uServe}.
% (Section~\ref{sec:tail}). 
Then, the  Chameleon scheduler combines this estimated output size with the known number of input tokens and the rank of the adapter required by the request to calculate a \emph{Weighted Request Size (WRS)} \circleint{2}. 

% \noindent\textbf{Scheduling Requests.}
Chameleon uses the WRS to categorize requests into classes, e.g., \emph{small}, \emph{medium},
and \emph{large}, and admit  them to different request queues  \circleint{3}.
Chameleon uses  iteration-level scheduling  (Section~\ref{sec:background}) and, therefore,  on every decode iteration, it can remove  and add  requests to a batch.
Importantly, on every iteration,
Chameleon tries to take  requests from all queues, while respecting the queues' assigned quotas. 
Specifically, each queue is assigned some amount of GPU resources that the requests from that queue can consume at each iteration.
\begin{comment}
These queues are assigned different amounts of GPU resources that their requests can consume at each iteration.
% I.e., 
\end{comment}
\begin{comment}
Chameleon uses different admission thresholds per queue to control the maximum number of requests admitted to a single batch from each queue. 
It sets them to be inversely proportional to the queue's associated size class and 
creates a fast lane for smaller requests while guaranteeing that no request will starve.
\end{comment}
% For maximum throughput, Chameleon assigns more resources for shorter requests, i.e., it 
The scheduler creates a fast lane for short requests preventing their HoL blocking,
but still admits requests from all queues
guaranteeing that no request will starve.
To handle load fluctuations and changes in request properties,
Chameleon dynamically adjusts both the   number of queues and the per-queue cutoffs based on the monitored WRS distribution of the incoming load.

% \noindent\textbf{Managing the adapter cache.}
On every scheduling decision, Chameleon also invokes its {\em Cache Manager}.
This is a software controller that manages the Chameleon Adapter Cache. Its function is to: (i) (pre)fetch any necessary adapters required by the requests to be scheduled and (ii) evict any idling adapters when the   GPU memory does not have room to store all the necessary state for the incoming requests \circleint{4}.
%' input, output and KV cache entries ~\circleint{4}. 
The Cache Manager tracks all cached adapters and the necessary metadata to enforce a  cost-aware eviction policy.

\vspace{-1mm}
\subsection{Chameleon Adapter Cache}
\label{sec:cache}

The Chameleon Adapter Cache (or Chameleon Cache, for short) is a software structure that stores unused adapters  loaded by previous requests  in idling GPU memory. The goal is to eliminate the costs of fetching the adapters again in the future, on the critical path of an inference request. 
Chameleon maintains one cache instance per LLM instance, i.e., each LLM replica has its own local adapter cache.
Each cache entry contains the adapter's weights and some metadata used for cache management. 
The metadata contains:

\begin{itemize}[leftmargin=*]
\item \textbf{Adapter ID}: A unique identifier for the adapter. 
\item \textbf{Adapter Rank}: The size of the adapter, which affects the amount of GPU memory it occupies. 
\item \textbf{Last Used Timestamp}: The last time the adapter was accessed.
\item \textbf{Usage Frequency}: The total number of times the adapter has been used within a specific time frame. 
\item \textbf{Reference Counter (RC)}: The number of active requests using this adapter. If RC is zero, the adapter is eligible for eviction.
\end{itemize}

The cache is managed by a Cache Manager.
The manager  performs the following operations: (i) it retrieves a cached adapter for an incoming request, (ii) it loads a missing adapter from host memory, (iii) it dynamically resizes the cache based on the incoming load, and (iv) it employs a cost-aware eviction policy to discard cached adapters when necessary.
Since adapter  weights are read only, there is no need to be concerned about their coherence, or to write them back to the host memory on eviction from the cache. Next, we describe three aspects of the Chameleon Cache.

\myparagraph{1. Dynamic Cache Sizing} Unlike in hardware caches, the capacity of the
Chameleon Cache changes dynamically over time. 
The Cache Manager adjusts its size in real time,  to 
ensure that the resource demands of incoming requests are always met
%.while consuming the fluctuating idle memory for adapter caching 
(Figure~\ref{fig:mem-time}).
For example, when the input activations, KV entries, or missing adapters of incoming requests
would not  fit in the available free GPU memory, the Cache Manager downsizes the Chameleon Cache, evicting adapters to free up the necessary space. Similarly, when a request ends and there is enough idling GPU memory, the Cache Manager expands the Chameleon Cache to store the request's  adapter. 

The Chameleon Cache Manager and Scheduler work synergistically for cache resizing. 
As the Scheduler scans the request queues and assembles a batch of requests to be scheduled on every decode iteration (Section~\ref{sec:scheduler}), it monitors the requests' memory requirements.
It then communicates the exact amount of memory required by the batch to the Cache Manager. The latter,
if necessary, discards unused adapters to free up space, based on a cost-aware eviction policy, and loads any missing adapters from the CPU memory. 

%The Scheduler monitors the incoming requests' memory requirements and invokes the Cache Manager. Specifically, it scans the request queues to assemble a batch of requests to be scheduled on every decode iteration (Section~\ref{sec:scheduler}) and communicates the exact amount of memory required by the batch to the Cache Manager. If necessary, the manager then discards unused adapters to free up space, based on a cost-aware eviction policy, and loads any missing adapters from the CPU memory. 

\myparagraph{2. Cost-Aware Eviction Policy}
The cache eviction policy in our multi-adapter environment needs to be more sophisticated
than existing policies based on recency like  least recently used (LRU). 
While these policies can capture the temporal locality in the adapter requests, they may fail to 
capture the adapters' skewed popularity found in LLM serving workloads. Indeed,
certain adapters are accessed more frequently than others, or are used by a larger number of concurrent inference requests~\cite{slora}. Evicting frequently-used adapters can increase miss rates and the CPU-GPU link bandwidth consumption due to frequent adapter reloading. 
%Indeed, in Section ~\ref{sec:eval} we show that least recently used (LRU) 
%policy alone fails to achieve optimal performance. 

An additional reason why LRU is insufficient is that cache misses in this environment have varying costs. Unlike hardware caches, which store cache lines of fixed size, the Chameleon Cache caches adapters,
which have different sizes, i.e., ranks. Consequently, the latency to load an adapter on a cache miss is not fixed. Larger adapters take longer to transfer from host to GPU memory. 
Thus, the eviction policy must be  \emph{cost-aware}~\cite{faascache} and prioritize the eviction of smaller adapters.

\begin{comment}
    Selecting an appropriate eviction policy is important for the cache performance,
and such a policy has to consider
the special characteristics of the multi-adapter environment. For example, while policies based on recency can capture the temporal locality in adapter requests, they may fail to cover the adapters skewed popularity found in LLM serving workloads, i.e. the fact that certain adapters are being accessed more frequently than others (used by a larger number of concurrent inference requests)~\cite{slora}. Evicting frequently used adapters can increase miss rates and CPU-GPU link bandwidth consumption due to frequent adapter reloading. 
Indeed, in Section ~\ref{sec:eval} we show that least recently used (LRU) 
policy alone fails to achieve optimal performance. 

An additional reason why LRU is insufficient on its own, is that cache misses in this environment have varying costs. Unlike hardware caches, that store objects with uniform sizes (cache lines), Chameleon caches objects, i.e. adapters, with different sizes, i.e. ranks. Consequently, the latency to load an adapter on a cache miss is not fixed, i.e. larger adapters take longer to transfer from host to GPU memory. Thus, the eviction policy must be also \emph{cost-aware}~\cite{faascache}, i.e. prioritize the eviction of smaller adapters.
\end{comment}

Similar to caching schemes employed in other domains~\cite{faascache}, we show that relying solely on a single feature is insufficient to capture the complex trade-offs in our system. 
To address this limitation, we propose a \textit{compound eviction algorithm} that 
%simultaneously 
considers multiple factors influencing adapter importance. Our scheme 
calculates a \emph{score} for each adapter based on its frequency of use, recency of access, and size. 
Frequency of use is significant, as some adapters are more popular than others. Recency is important for temporal locality, as bursts of requests for the same model/adapter are a common access pattern~\cite{splitwise}. The size of the adapter affects the cost of a cache miss, as  larger adapters are costlier to reload. We combine these factors linearly to calculate an eviction score:
%The score is computed using the formula:
$Score = F \times Frequency + R \times Recency + S \times  Size$,
where  F, R, and S are weighting coefficients.
The adapter with the lowest score is considered the least critical and is evicted first. 
%This approach balances the trade-offs between evicting adapters that are infrequently used, not recently accessed, or large in size. 

The weighting coefficients 
F, R, and S enable adjusting the sensitivity of the eviction policy to each factor.
For example, if frequency is deemed more 
%indicative of future usage patterns 
important than recency for the running workload, F can be assigned a higher value relative to 
R. For this study, we use static coefficients, which  we set by offline profiling of industrial traces of inference requests~\cite{splitwise} combined with adapter size distributions found in the literature~\cite{slora}. F, R, and S are set to 0.45, 0.10, and 0.45, respectively. 
%\hl{We have also measured that even a simplistic online dynamic %adjustment of the coefficients is adequate, with
%less than 5\% performance drop.}
%We \hl{study \system{}'s sensitivity to this tuning in} \S~\ref{sec:newsensitivity} \hl{and} consider %an 
%\hl{their}
%online dynamic adjustment %of the coefficients
%based on the incoming load for future work. 

Chameleon never evicts adapters that are actively used by running requests. To guarantee this, the Cache Manager maintains a reference counter per adapter that tells how many requests are currently using the 
adapter. It  considers eligible for eviction only the adapters whose counters have dropped to zero. Additionally, the Manager checks the Scheduler to identify adapters associated with queued requests.
Such requests are  not currently running,  but are guaranteed to execute in the near future. The Manager attempts to retain these adapters in the cache, provided there is sufficient memory available. 
%However, if memory constraints arise, adapters for queued requests may be evicted to accommodate the running batch.
Overall, the Manager applies the eviction policy only to adapters that are not used by both currently-running and queued requests. The adapters of queued requests are considered for eviction only when memory constraints make it necessary.

\myparagraph{3. Prefetching} Building on prior-art optimizations, Chameleon monitors the request queues and,
whenever possible, prefetches  the missing adapters required by the waiting requests before they are admitted to a batch for execution~\cite{slora,dlora-osdi}.  This approach can lead to late prefetching, where the request becomes ready for execution before the prefetching is completed.
Hence, we explore techniques that predict future load, such as a histogram-based approach~\cite{serverless_wild}, and prefetch adapters even for requests that are not currently queued. 
Since the effectiveness of this optimization depends heavily on the prediction accuracy, we do not enable it by default in our evaluation;  we include a separate experiment to assess the potential impact of such 
prefetching. % (Figure~\ref{fig:prefetch}).

Prefetching has the potential to hide the cost  of 
loading an adapter's weights and remove it from the request's critical path of execution.
However, prefetching 
%and unloading the adapters on completion 
still consumes CPU-GPU link bandwidth. Thus, caching the adapters is still necessary for high performance.

%\myparagraph{\hl{4. Multi-GPU set-up.}}

%\noindent\hl{\textbf{Proactive vs reactive eviction.}}\todo{Add here content.}

\subsection{Chameleon Scheduler}
\label{sec:scheduler}

The \system{} Scheduler is
inspired by prior work on load balancing multi-server environments with heterogeneous task size distributions~\cite{qzilla, sita, sita2}. 
The scheduler stores the inference requests across multiple queue lanes, each dedicated to handling requests within a specific size range. 
Its goal is twofold: to provide a fast lane for small requests, preventing their HoL blocking, and to ensure that requests from all lanes are scheduled in parallel, avoiding starvation for large requests.

Each queue is assigned a resource quota, which governs the resources available to execute   requests from that queue. This quota is represented as tokens, and includes input tokens, output tokens, and tokens due to
the memory required for the corresponding adapter. These tokens determine the resources a queue can reserve for request execution. 
When a request from a specific queue is admitted to the batch, the queue's available quota is decreased by the 
memory consumption of the request, determined by its input and output length, and adapter size,
all translated into tokens.
%the request's memory consumption, subtracting the memory required for its input and output, its adapter and its KV cache, all translated into tokens.
When the request ends, it returns the borrowed quota back to the queue. In every iteration, all queues have the chance to put requests into the batch, although the queues with smaller requests are accessed first.

%hence we will refer to them as "higher" request queues. 

\vspace{1pt}
\noindent \textbf{1. Admission to the Queues.} We characterize the requests entering the system by three parameters: known input size, predicted output size, and rank of the used adapter. 
We calculate the weighted request size (WRS) as an estimate of the total execution time of a request 
based on  the formula: \[
\text{WRS} = \left(A \cdot \frac{\text{InputSize}}{\text{MaxInputSize}} + B \cdot \frac{\text{OutputSize}}{\text{MaxOutputSize}} \right) \cdot \frac{\text{AdapterSize}}{\text{MaxAdapterSize}}
\]
The input size affects the prefill latency, which is typically shorter than the decode latency but  still a significant contributor to the total execution. 
The  output size determines the number of decode iterations, which affects the decode latency and 
the total execution time.
The adapter size affects the speed of both prefill and decode processes (Section~\ref{sec:background}).
It can be shown that using this polynomial of degree 2 
improves Chameleon's performance by up to 10\% over using a polynomial of degree 1 that simply
combines the three factors linearly.

\begin{comment}
    The   input size affects prefill execution, the  output size determines the number of decode iterations, and the adapter size affects the speed of both prefill and decode processes.
It can be shown that using this polynomial of degree 2 
improves Chameleon's performance by up to 10\% over using a polynomial of degree 1 that simply
combines the three factors linearly.
\end{comment}

%linear formula to combine the three different factors. 
%We propose this WRS nonlinear polynomial expression to capture the correlation between these factors---e.g., how the adapter size correlates with the input length affecting prefill performance and with the output length affecting decode latency. It can be shown that using a polynomial formula improves Chameleon's performance by up to 10\% over using a linear formula to combine the three different factors. 

We call  WRS 
%estimated total execution time as 
the "request size", and use it to classify requests into size ranges and dispatch them to 
corresponding queues for scheduling. 
%As discussed next, \system{} promotes the admission of smaller requests per batch, i.e. requests that run faster, targeting the reduction of head of line blocking. \system{} does not categorize and promote/demote requests based on their memory footprint, i.e., WRS does not quantify requests in terms of memory consumption. However, we find that requests that run longer typically have larger footprints also.
{\em A} and {\em B} are weighting coefficients  chosen based on our
sensitivity studies and on profiling in Section \ref{sec:char}. We set {\em A} to 0.4 and {\em B} to 0.6.

%\vspace{-3mm}
%For example, 
%The input size affects the prefill latency, which runs typically faster than decode but is still significant to total execution. 
%The output size dictates decode latency, and thus is expected to significantly affect %primarily determines 
%the total execution time. It is predictive and, hence, assigning its coefficient to a too high value would make the system fragile to prediction accuracy.
%\hl{Similarly, the input size affects the prefill latency, which runs typically much faster than decode but is still significant to total execution. 
%Adapter size affects both prefill and decode (\S~\ref{sec:background}).
%In fact, in our experiments} \S~\ref{sec:results} \hl{we found that a linear and a non-linear WRS estimation have the same accuracy.}
%Prefill operations on the GPU are completed much faster than decode operations, and adapter size affects the speed of both prefill and decode processes.
%\vspace{2mm}
Given a request, the scheduler uses the calculated WRS and the per-queue cut-offs (i.e., the boundaries that define the ranges of request sizes for each queue) to
place  the request in the correct queue. 
Later, we detail how to determine these per-queue cut-offs using request clustering.
Note that the Chameleon Scheduler uses an open-source BERT-based proxy model to predict a request's output length~\cite{uServe}.

%\vspace{-1mm}
\begin{algorithm}[t]
\small
\SetAlgoLined
\SetKwProg{Def}{def}{:}{}
\Def{generate\_batch}{
\SetKwData{Input}
\KwData{\textbf{Inputs:} Queues = requ. queues; PQ\_Tokens = per queue tokens\\}
\KwResult{Batch of requests to be sent to the GPU.}
\SetKw{Break}{break}
\vspace{1pt}

batch $\gets$ []\;
leftover $\gets$ 0\;
\For(\tcp*[h]{Phase 1}){each q in Queues}{
consumed $\gets$ \texttt{put\_batch}(q, PQ\_Tokens[q], batch)\;
\If{q is empty}{
leftover $\gets$ leftover + (PQ\_Tokens[q] - consumed)\;
}
}
\For(\tcp*[h]{Phase 2}){each q in Queues}{
\If{leftover == 0}{
\Break\;
}
consumed $\gets$ \texttt{put\_batch}(q, leftover, batch)\;
leftover $\gets$ leftover - consumed\;
}
\KwRet{batch}\;
}

\vspace{5pt}

\SetKw{Continue}{continue}
\SetKw{Break}{break}
\SetKw{Append}{append}
\SetKwProg{Def}{def}{:}{}
\Def{put\_batch}{
\SetKwData{Input}
\KwData{\textbf{Inputs:} Queue; Tokens; Batch\\}
\KwResult{Tokens consumed by added requests from the queue.}
\vspace{1pt}

resources $\gets$ Tokens\;
consumed $\gets$ 0\;
\For{each req in Queue}{
needed $\gets$ \texttt{need\_resources(req)}\;
\If{resources $<$ needed}{
\Break\;
}
resources $\gets$ resources - needed\;
consumed $\gets$ consumed + needed\;
batch.append(req)\;
}
queue $\gets$ [req \textbf{for each} req \textbf{in} queue \textbf{if} req \textbf{not in} batch]
\KwRet{consumed}\;
}
\caption{Generate a new batch of requests.}
\label{alg:generate_batch}
\end{algorithm}
\myparagraph{2. Admission to the Batch} The idea behind the \system{} Scheduler 
is depicted in Algorithm \ref{alg:generate_batch}. It operates in two  phases: \emph{Initial Request Admission} and \emph{Redistribution of Spare Resources}. In the first phase, each queue attempts to put requests into the batch, up to the queue's maximum allowed resources. If certain queues have few or no requests to put, any unused resources are collected and consolidated into a {\em Total Spare Resources} bucket.
After the first phase ends, in the second phase, the scheduler redistributes the spare resources to queues 
that still have pending requests, aiming to maximize resource utilization. Specifically, starting from the smallest-request queue and moving downward to larger-request queues, the scheduler allocates as much of the spare resources as possible to admit waiting requests into the batch. If requests from a given queue still cannot be admitted due to insufficient tokens available, no additional resources are allocated to that queue.

\begin{figure}[t]
\centering 
\includegraphics[width=\columnwidth]{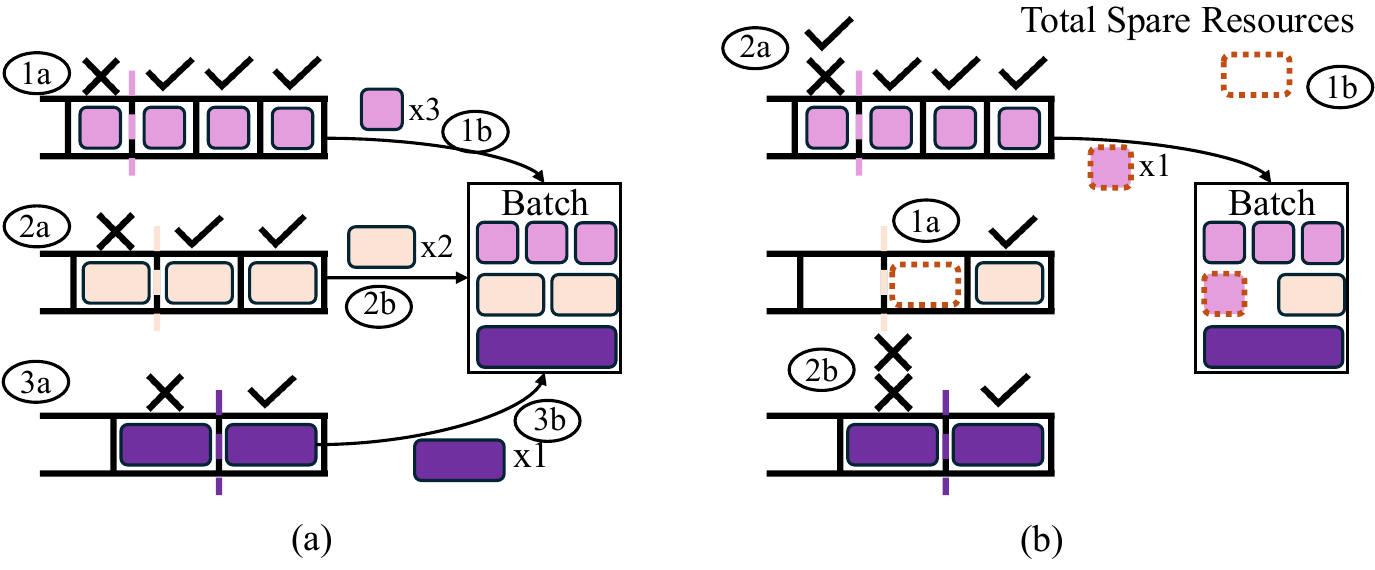} 
\vspace{-9mm} 
\caption{An example of Chameleon Scheduler operation when (a) no spare resources are collected (b) spare resources are collected and redistributed.} 
\label{fig:chameleon-sched} 
\vspace{-6mm} 
\end{figure}

The phases of this process are illustrated in Figure \ref{fig:chameleon-sched}, which depicts three request queues, for "small", "medium", and "large" requests. 
Figure \ref{fig:chameleon-sched}(a) shows the case
when no spare resources are collected, while Figure \ref{fig:chameleon-sched}(b) 
shows the case when spare resources are collected and redistributed.
In Figure \ref{fig:chameleon-sched}(a), the  Initial Request Admission phase 
starts with the small-request queue, admitting three requests, which fit within the queue's resource quota  \circleint{1a}. 
The fourth request is not admitted due to insufficient  resources allocated to the queue. 
These admitted requests are then placed into the batch \circleint{1b}. 
The same procedure is subsequently applied to the medium-request queue \circleint{2}, and the large-request queue \circleint{3}. 
At the end of this phase, there are no remaining resources to redistribute, so the process concludes without entering the second phase.

%\system{} never reserves %idle 
%resources per queue in anticipation of future events, e.g. when a queue is empty. Instead, it collects each queue's spare tokens and re-distributes them top-down, i.e. starting from the queue that hosts small requests. 
Figure \ref{fig:chameleon-sched}(b) shows a case where spare resources are available for redistribution. 
Specifically, the medium-request queue only has a single request, and so it does not use up its allocated resources. Hence, during the Initial Request Admission phase,
the queue contributes with some
spare resources \circleint{1a}, which are deposited into
 the Total Spare Resources bucket \circleint{1b}. 

Then, during the Redistribution of Spare Resources  phase, the scheduler checks each queue 
in order, from the small- to the medium- and large-request queue, to try to admit any remaining requests into the batch.
The small-request queue evaluates if its single pending request can be admitted \circleint{2a}. 
Since the Total Spare Resources are sufficient, the scheduler allows this request  to be admitted.
The medium-request queue
has no pending requests and is skipped. 
The large-request queue  attempts to put its pending request \circleint{2b}. 
However, since the remaining spare resources are not enough,
%the available resources are insufficient, 
the request remains in the queue. At the conclusion of this phase, the batch is finalized and ready for execution.
%\hl{Explain WRS outliers/false categirization, can a request with large input and adapter but small output land on the wrong queue?}

\noindent \textbf{3. Opportunistic Bypassing.} 
Sometimes, a request {\em R1} that should be put in a batch according to the resource quota allocated to  
its queue, may fail to get
admitted to the batch because  there is not enough idle GPU memory to store its adapter---even after \system{} evicts all idle cached adapters. In such case, without proper action, the queue is unable to use its
allocated resource quota.
However, it may happen that a younger request {\em R2} in the same queue uses an adapter that is 
either already loaded in the Chameleon Adapter Cache or  is small enough to fit in the remaining space of
the cache.

To address this case, Chameleon implements an {\em Opportunistic Bypass} mechanism, whereby 
{\em R2} is put in the batch for execution,  bypassing {\em R1}.
This mechanism improves system throughput by allowing more requests to be processed without waiting for cache space to become available.
However, repeated bypassing can lead to request starvation.
Consequently, Chameleon first predicts how soon will the memory needed by {\em R1} become available,
and how long will {\em R2}'s execution take. Then, Chameleon allows {\em R2} to bypass {\em R1} only
if the former is longer than the latter.
%a request to bypass the current head of the queue, only if the 
%admitted request will 
%not exceed the waiting time for the request at the head of the queue. 
%We rely on the prediction of request length when deciding whether to allow adapter bypassing.

Unfortunately, predictions may turn  out to be  wrong. Hence, if, before {\em R2}'s execution completes, 
Chameleon finds enough free memory on the GPU (including the memory used by {\em R2}) to execute {\em R1},
it squashes {\em R2} for later re-execution.
In our experiments, we see at most 5\% of requests getting squashed. 
Note that these scenarios can only happen when the GPU memory is entirely consumed by running requests and no idle adapters are cached. 
%they are evicted to free up space for the blocking request. 
Thus, eviction policies do not apply here.

\myparagraph{4. Determining the Number of Queues} 
The efficiency of the Chameleon %\system{}
Scheduler depends on the number of used queues.
Too few queues may cause HoL blocking when there is a high variability in the request sizes within a queue, while too many queues can result in load imbalance and underutilized queue resources 
%when requests are homogeneous 
due to  resource fragmentation.
To decide on the optimal number of queues, the
Chameleon Scheduler uses %a method based on 
\emph{K-Means clustering}.
Given the distribution of request sizes,
the scheduler computes K-Means clustering for values of K ranging from 1 to $K_{max}$.
With K-Means clustering,
requests similar in size are grouped within the same cluster, and requests from different clusters are different enough to require separate resources.
For each value of K,
the scheduler calculates
the Within-Cluster Sum of Squares (WCSS), and picks the K that yields minimal WCSS as the optimal number of queues. 
We set the maximum number of queues, $K_{max}$,
to 4 to keep queue management overheads tolerable.

Once we have the $K$ centroids from the clustering result, we proceed to 
determine the per-queue request-size cutoffs.
Specifically, we define the cluster boundaries as the midpoint between the centroids of two consecutive clusters.
For example, the boundary between $Cluster_i$ and $Cluster_{i+1}$ is ($Centroid_{i} + Centroid_{i+1}$)/2.
The boundaries 
represent the maximum and minimum request sizes for each queue:
$Queue_1$ handles requests smaller than $Boundary_1$,
$Queue_2$ handles requests larger than or equal to $Boundary_1$ but smaller than $Boundary_2$, and so on for
all $K$ queues.

\begin{comment}
Each request size is assigned to one of the 
$K$ clusters based on its proximity to the cluster centroids.
To determine the per-queue cutoffs, we exploit the results of K-Means clustering.
We obtain the $K$ centroids from the clustering result,
and define the cluster boundaries as the midpoint between the centroids of two consecutive clusters.
For example, the boundary between $Cluster_i$ and $Cluster_{i+1}$ is $\frac{Centroid_{i} + Centroid_{i+1}}{2}$.
The boundaries 
represent the maximum request size for each queue:
$Queue_1$ handles requests smaller than $Boundary_1$,
$Queue_2$ handles requests larger than $Boundary_1$ but smaller than $Boundary_2$, and this continues to all $K$ queues.
\end{comment}

The distribution of request sizes changes over time due to  fluctuating load behavior.
Hence, static queue configurations can lead to inefficiencies.
Therefore, \system{} dynamically adjusts the number of queues based on the observed load patterns.
Specifically, the system periodically gathers recent request data to analyze the distribution of request sizes
and, every $T_{refresh}$,  re-computes the optimal number of queues and the per-queue cut-offs using the aforementioned method.
Since changes in load patterns are not sharp~\cite{dynamollm}, changing the multi-queue organization 
happens relatively infrequently. We set
$T_{refresh}$ to 5 minutes, which
adds negligible overheads.

%$T_{refresh}$ can be 5 minutes, which induces negligible overheads.

\myparagraph{5. Assigning Quotas per Queue} After determining the number of queues in the system and the per-queue cut-offs, the
Chameleon Scheduler  assigns the resource quotas to each queue.
%These quotas govern how many resources can be assigned for requests from a given queue.
For this, we use queuing theory, modeling the system as $K * M/M/1$ queues~\cite{usteal}.

We take the maximum allowed size ($S$) of a request in a queue in tokens, 
the assigned resource quota ($Tok$) to the queue  in tokens, the expected time duration ($D$) of processing a request from the queue, the arrival rate ($\lambda$) of requests to the queue, and the requests' $SLO$.
Then, the processing rate of the requests is $\mu = \frac{Tok}{S * D}$, while the total time that a request spends in the system is $T_{total} = \frac{1}{\mu - \lambda}$.
To meet the $SLO$,
the system needs to satisfy the following equation: $T_{total} \leq SLO$.
Combining these constraints, we compute the minimum assigned quota
in tokens ($Tok_{min}$) to the queue that is required for requests from the queue
to meet the SLO: $$Tok_{min} \geq S * D * \left(\frac{1}{SLO} + \lambda\right)$$ 

The total number of available tokens in the system ($Tok_{total}$) must be greater than or equal to $\sum\limits_{q} Tok_{min}^q$ (i.e., the sum of the minimum number of tokens needed by each queue $q$).
Then, each queue $q$ is assigned its minimum number of required tokens ($Tok_{min}^q$), and the remaining tokens ($Tok_{total}$ - $\sum\limits_{q} Tok_{min}^q$)  are split across queues proportionally to their initial weights.
%\hl{As explained earlier the quota tokens of every queue are spent for each request's KV cache, adapter and input and output tokens.}
%\hl{As explained, earlier the quota tokens of every queue are consumed by the request's }
%\myparagraph{\hl{6. Multi-GPU set-up.}}

To adjust to the dynamic nature of the workload,
Chameleon recomputes the per-queue  quotas every $T_{refresh}$.

\subsection{Multi-GPU Set-up}
\label{sec:multigpu}

With multiple GPUs, LLM inference can
use tensor parallelism (TP), pipeline parallelism (PP), and data parallelism (DP). 
In TP or PP, Chameleon distributes its adapter cache accordingly, so each GPU
stores a fraction of each adapter;
%(partitioned across tensors or layers); 
in DP, Chameleon  replicates the adapter cache across engines. Since adapters
are read-only, data coherence is not a concern. In this paper, we follow the S-LoRA TP strategy \cite{slora}.

For scheduling, in TP or
PP, 
Chameleon treats all GPUs as a single execution engine; in DP,
Chameleon  uses a two-level scheduler: a global scheduler
dispatches requests to the different engines, and each engine has its local scheduler.

\begin{comment}
LLM inference scales vertically within a node via model parallelism—tensor parallelism (TP), which shards parameters and compute across GPUs, and pipeline parallelism (PP), which partitions layers into stages—and horizontally across nodes via data parallelism (DP), which deploys multiple model serving instances to increase throughput. In TP configurations, Chameleon treats all GPUs within a node as a single execution engine: it shards each adapter across devices and distributes the adapter cache accordingly. We follow the S-LoRA \cite{slora} TP strategy. Under DP across multiple nodes, adapters are replicated in per-node caches; because adapters are read-only at inference time, no coherence protocol is required. Scheduling is two-level: a global dispatcher load-balances requests across engines, while a per-engine Chameleon scheduler batches and orders requests according to the requests' arrival and size patterns.
\end{comment}
%\section{Discussion}
%\label{sec:discussion}

%\noindent\textbf{\hl{6. Multi-GPU set-up}} \todo{Discuss data and tensor parallelism.}

\section{Evaluation}
\label{sec:eval}

\subsection{Evaluation Methodology}
\label{sec:methodology}

\noindent \textbf{Hardware Platforms and LLMs.}
We run most of our experiments on a server equipped with an A40 NVIDIA GPU~\cite{a40} and  an AMD EPYC 9454 CPU.
% (48 cores and 377GB of main memory).
The GPU has 48GB memory and the CPU has 48 cores and 377GB of main memory.
For the scalability experiments,
we use a server equipped with an A100 NVIDIA GPU~\cite{a100} configured with 24GB, 48GB, and 80GB of GPU memory. For the multi-GPU experiments, we use four A100 GPUs with 80GB of GPU memory. 
For the majority of experiments, we use the Llama-7B~\cite{llama2} model.
When memory capacity allows, %in addition to the Llama-7B model,
we also run the Llama-13B and Llama-30B models.
% While we present the results with Llama series of models,
We used other models, such as Falcon~\cite{falcon},
OPT~\cite{opt}, and Mixtral~\cite{mixtral2}
and observed similar trends.

\vspace{0pt}
\noindent \textbf{Workload Configuration.}
We set the input and output
lengths of requests based on the open-source production trace from Azure~\cite{splitwise}.
To vary the load on the system, we use the Poisson distribution for the request inter-arrival time~\cite{punica,caraserve,alpaserve}.
We set the number of different adapters used by the requests to $N_a$.
Unless specified otherwise, in our experiments, $N_a$ is 100. 
There are five adapter ranks:  8, 16, 32, 64, and 128.
Each rank has an equal number of different adapters, i.e., $N_a / 5$.
To each request, we attach an adapter, following a uniform distribution for rank popularity
and a power-law distribution 
for adapter popularity within a rank~\cite{slora}.

 %and a uniform distribution for rank popularity.

\begin{comment}
Then, for a request we pick the adapter rank following the power-law distribution, smaller adapters are more likely to be chosen.
After choosing the rank, we pick one of the adapters of a given rank with a uniform distribution.
\end{comment}

\vspace{0pt}
\noindent \textbf{Baseline Systems.}
We run the experiments on S-LoRA~\cite{slora}, an open-source state-of-the-art LLM inference serving platform for adapter environments, and compare \system{} to the baseline S-LoRA.
%  Note that 
S-LoRA performs iteration-level scheduling using a FIFO policy, and asynchronous adapter 
prefetching without adapter caching.
%\hl{We compare \system{}'s caching replacement policy to LRU and GDSF.}
We also compare \system{}'s scheduler to the recently proposed SJF scheduler in $\mu$Serve~\cite{uServe}.
We measure Time-To-First-Token (TTFT), Time-Between-Tokens (TBT), and End-To-End (E2E) latency.
%As in prior work, 
We set the SLO to be 5$\times$ the average request execution time in a low-load system~\cite{dynamollm,alpaserve,splitwise}.
% In the following, we show our main results.

\subsection{Performance Gains}
\label{sec:results}

\begin{figure}[t]
\centering
\includegraphics[width=0.9\columnwidth]{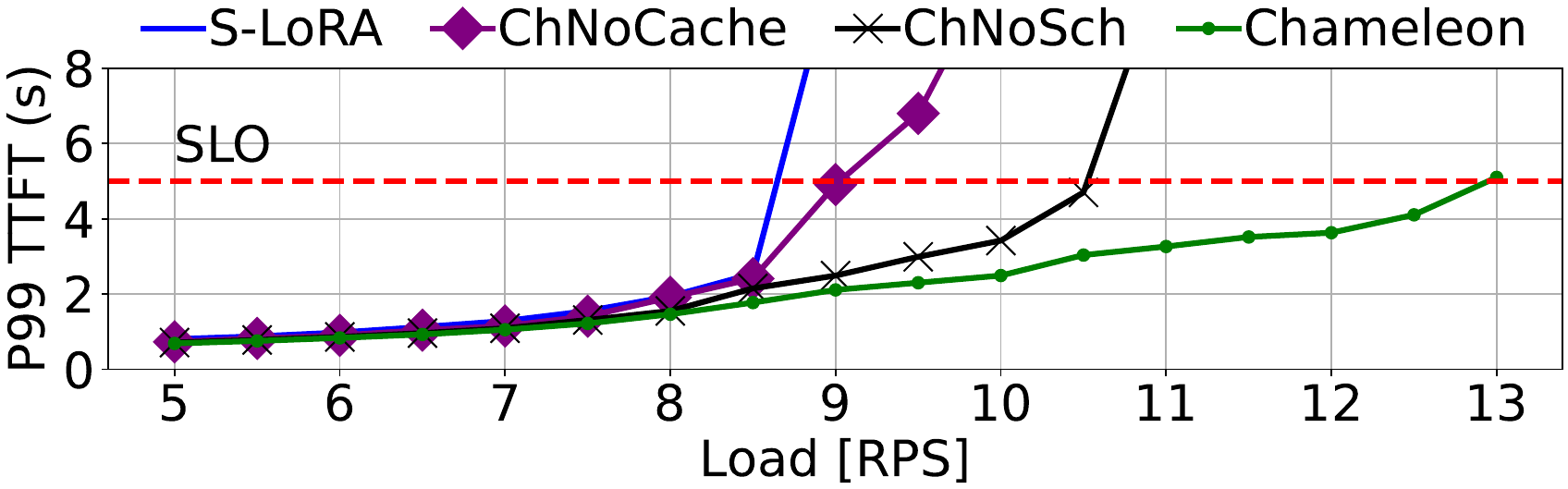}
\vspace{-4mm}
\caption{P99 TTFT tail latency for \emph{S-LoRA}, \emph{ChameleonNoCache}, \emph{ChameleonNoSched}, and \emph{\system{}} under different loads.
The red dashed line indicates the SLO.}
\label{fig:overall}
\vspace{-3mm}
\end{figure}

\vspace{0pt}
\noindent \textbf{1. Tail Latency.}
Figure~\ref{fig:overall} shows, for different loads, the P99 TTFT tail latency for \emph{S-LoRA},
\emph{\system{}} without its cache, \emph{\system{}} without its scheduler, and the full
\emph{\system{}}. We consider the latter in this section. Although it is hard to see for low RPS,
\emph{\system{}} consistently has lower TTFTs than \emph{S-LoRA}, 
and the benefits become more pronounced as the load increases.
At low (6 RPS), medium (8 RPS), and high (9 RPS) loads,
\emph{\system{}} reduces the TTFT tail latency over \emph{S-LoRA}
by 14.7\%, 24.6\%, and 80.7\%, respectively.

\begin{figure}[t]
\centering
\includegraphics[width=0.9\columnwidth]{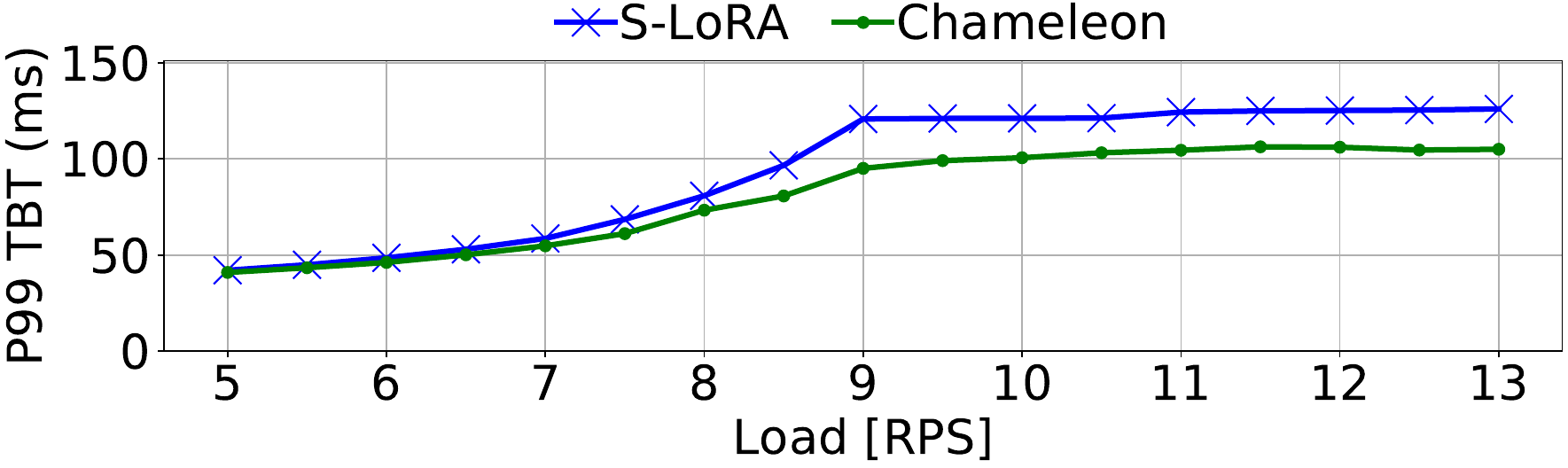}
\vspace{-3mm}
\caption{P99 TBT tail latency for \emph{S-LoRA} and \emph{\system{}}.} %  under different loads.}
\label{fig:tbt}
\vspace{-5mm}
\end{figure}

There are two reasons why \emph{\system{}} reduces the P99 TTFT latency over \emph{S-LoRA}.
First, its caching mechanism reduces the adapter fetching time and, as the load increases, it also alleviates the PCIe bandwidth bottleneck---which in turn further decreases TTFT latency.
%(Figure~\ref{fig:adapter-pcie-new}).
Second, its scheduling policy reduces  queueing delays, as it removes HoL blocking and prevents starvation, especially helping the requests at the tail.
As the load increases, GPU memory is increasingly consumed by the KV cache entries of the running requests, and there is less space for Chameleon to cache adapters not currently in use. It can be shown that, by 12.5 RPS, most of the time, GPU memory is fully used and there is no space for caching
adapters not in use. Still, we see that \system{}  manages to reach this point while keeping TTFT under SLO. Below 12.5 RPS, \system{} judiciously re-purposes scarce idle memory, caching
frequently-used and costly to reload adapters, while prioritizing requests with short execution times that use them. S-LoRA, on the other hand, already violates SLO at about 8.5 RPS, well before it can fully utilize all the available GPU memory to run requests.

\emph{\system{}} reduces both TTFT and TBT tail latencies.
Figure~\ref{fig:tbt} shows the P99 TBT latency for \emph{S-LoRA} and \emph{\system{}} under different loads. Again, \emph{\system{}} has lower latencies than \emph{S-LoRA} for all loads.
However, both systems keep their TBT latency under the SLO (150ms).
The reason is that TBT latencies are less affected by  queuing effects, and requests do not wait on adapter loading.
%\noindent\textbf{3. TBT and base model overloading.}
Substantially increasing the batch size can in theory increase TBT in both systems.
%, due to costlier operations over the base model. %\hl{
%For that, \system{} has a max\_batch\_size threshold, i.e. a maximum number of tokens that can be admitted to a batch. When reached, \system{} stops admitting requests even if there is enough available GPU memory for their KV caches and adapters. 
However, in our experiments of Figure~\ref{fig:tbt},  
%though,  this threshold is never reached, 
%i.e. 
we find that admissions %to batches 
are eventually
limited by  available GPU memory,
and batches never grow to the extent of violating TBT SLOs.
%and TBT remains under SLO without any dedicated admission policy.} 
%\XA{Should we position it like this, that we have such a threshold or that we should have one? What would be the number of tokens for it?} 

\begin{figure}[t]
\vspace{-1mm}
\centering
\includegraphics[width=0.9\columnwidth]{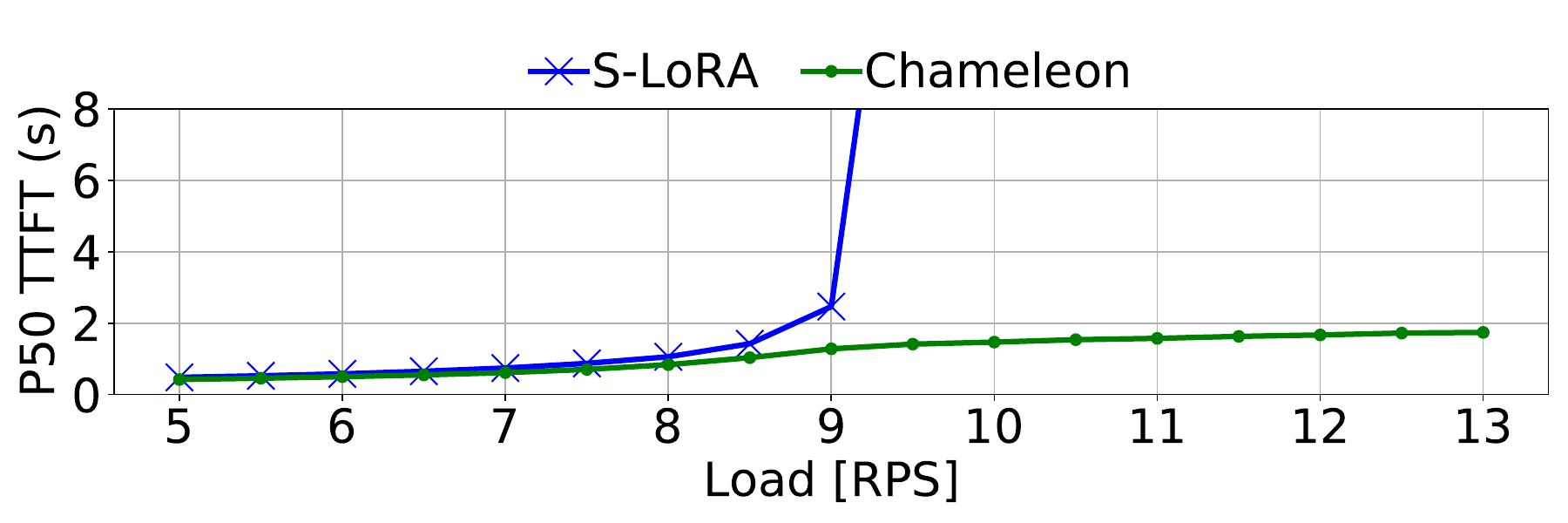}
\vspace{-4mm}
\caption{P50 TTFT latency for \emph{S-LoRA} and \emph{\system{}}.} %  under different loads.}
\label{fig:avg}
\vspace{-4mm}
\end{figure}

\vspace{0pt}
\noindent \textbf{2. Throughput.}
Figure~\ref{fig:overall} also shows the expected TTFT SLO as a red dashed line (i.e.,
5$\times$ the average request execution time in a low-load system).
We define the throughput as the load that a system can sustain without violating this SLO.
From Figure~\ref{fig:overall}, we can see 
that \emph{S-LoRA}'s starts violating the SLO around 8.6 RPS, 
while \emph{\system{}}'s starts violating the SLO around 12.9 RPS.
This 
results in 1.5$\times$ higher throughput for \emph{\system{}}.

\vspace{0pt}
\noindent \textbf{3. Median Latency.}
Figure~\ref{fig:avg} shows the P50 TTFT latency for \emph{S-LoRA} and \emph{\system{}}
under different loads.
At low (6  RPS), medium (8 RPS), and
high (9 RPS) loads,
\emph{\system{}} reduces the median latency over \emph{S-LoRA}
by 13.9\%, 20.9\%, and 48.1\%, respectively.
The benefits of \emph{\system{}} are still significant, although not as pronounced as in the tail latency.
The reason for this is that  average conditions are less demanding.

%on average, even in \emph{S-LoRA}, requests do not experience HoL blocking and adapter loading does not happen in bursts.

\vspace{0pt}
\noindent \textbf{4. Performance Breakdown.}
To understand the performance benefits of the two main \emph{\system{}}  techniques, we run them in isolation.
Figure~\ref{fig:overall} shows the P99 TTFT latency of \emph{Chameleon} when running only with either 
our caching technique (\emph{ChameleonNoSched}) or our scheduling technique (\emph{ChameleonNoCache}). 
We see that both systems improve the throughput over  \emph{S-LoRA}:
\emph{ChameleonNoSched} and \emph{ChameleonNoCache} have
1.2$\times$ and 1.05$\times$ higher throughput, respectively, than \emph{S-LoRA}.
However, their throughput is substantially lower than  \emph{\system{}}'s.
Hence, both adapter caching and adapter-aware scheduling are needed.
%we need to synergistically perform both adapter caching and adapter-aware scheduling.

\begin{figure}[t]
\vspace{-1mm}
\centering
\includegraphics[width=0.9\columnwidth]{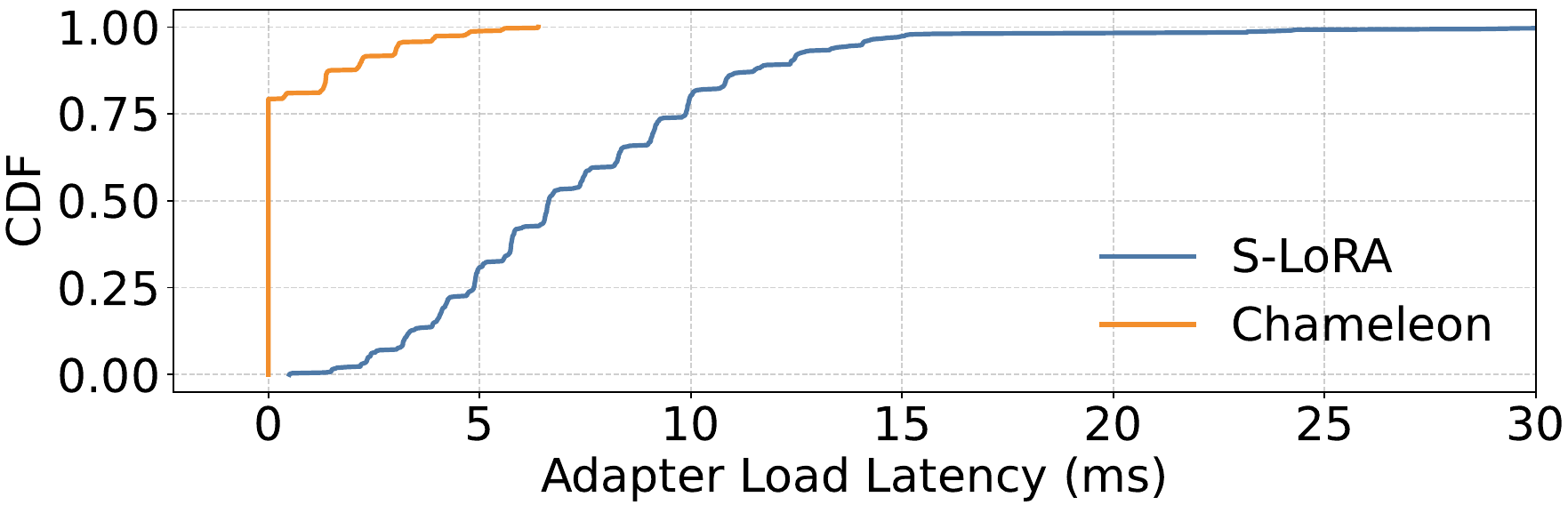}
\vspace{-4mm}
\caption{CDF of adapter loading latency on the critical path.}
% for \emph{S-LoRA} and \emph{Chameleon}.}} %  under different loads.}
\label{fig:ad-load}
\vspace{-9mm}
\end{figure}

\noindent 
\textbf{5. Adapter Loading Time.}
%This section studies how the Chameleon Cache reduces adapter load overheads compared to S-LoRA. 
Figure~\ref{fig:ad-load} shows the CDF of the latency of adapter loading on the critical
path for the requests of the Splitwise trace~\cite{splitwise} in \emph{\system{}} and
\emph{S-LoRA}. \emph{S-LoRA} suffers from adapter loading latencies of up to 30ms, as its prefetching scheme fails to completely overlap adapter transfer with computation. With \emph{\system}, on the other hand, 75\% of the requests hit in the Chameleon Cache, resulting in zero loading overheads, while the remaining 25\% of the requests pay loading costs of only up to 6ms. Adapter loading in \emph{\system{}} is cheaper because:
a) \emph{\system{}} prioritizes the eviction of smaller adapters and thus reloading on a cache miss is cheaper, and b) \emph{\system}'s caching reduces the contention on the PCIe.

%\noindent 
%\hl{\textbf{6. Memory Usage.}
%This section measures the memory usage with Chameleon as a function of time.
%It is shown in} Figure~\ref{}\hl{. We see that, compared to} Figure~\ref{fig:mem-time}\hl{, Chameleon is able to
%utilize most of the memory... }
 
\vspace{2mm}
\subsection{Different Scheduling/Caching Policies}
\label{sec:scpolicies}

In earlier experiments, we  compared Chameleon with  S-LoRA, which performs FIFO request scheduling and does not cache unused adapters.
Here, we compare Chameleon to a 
SJF (shortest-job-first) scheduling policy proposed by $\mu$Serve~\cite{uServe}.
Also, we augment the baseline with a Chameleon Cache that uses an LRU eviction policy.
%Also, we augment the baseline with Chameleon Cache using i) a simple LRU \hl{and ii) Greedy Dual Size Frequency (GDSF)} eviction policies
%\hl{to study the efficiency of Chameleon's adapter-aware evictions.}

\begin{figure}[t]
\centering
\includegraphics[width=0.9\columnwidth]{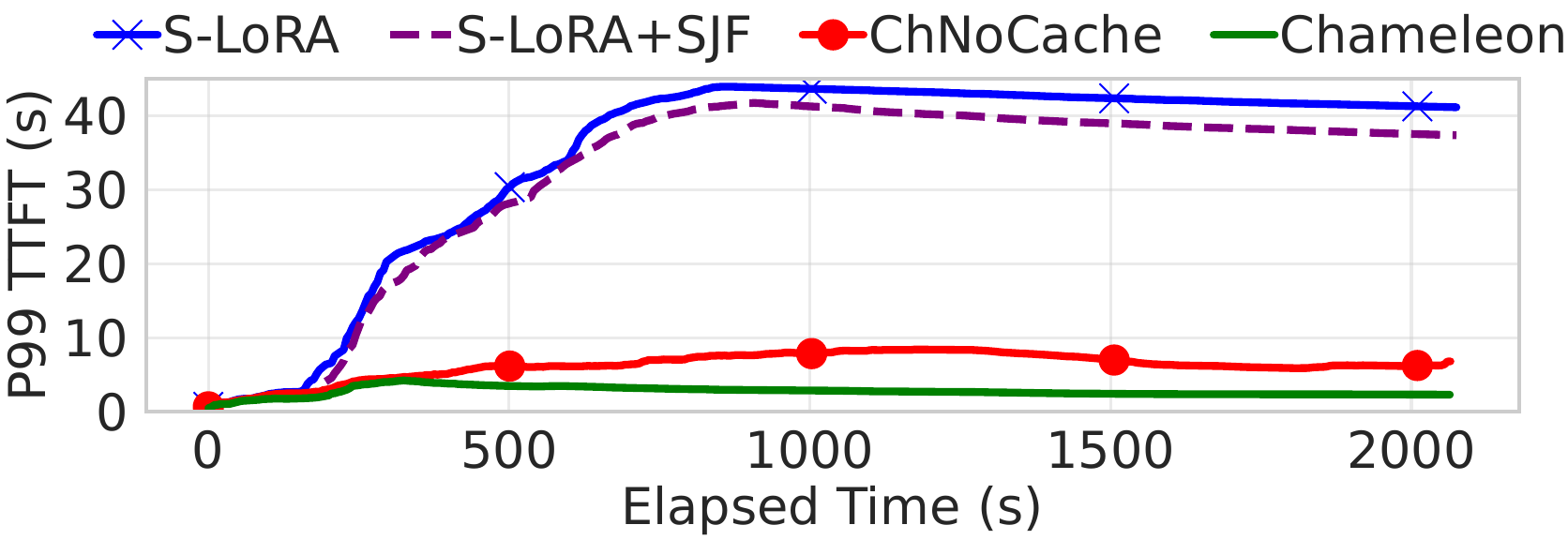}
\vspace{-4mm}
\caption{P99 TTFT latency over time with different scheduling policies: FIFO (default in \emph{S-LoRA}), SJF in \emph{S-LoRA}, and our proposed policy in \emph{ChameleonNoCache} and \emph{\system{}}.}
%Load is driven by the production trace~\cite{splitwise} at 9 RPS.}
\label{fig:schedres}
\vspace{-2mm}
\end{figure}

%\emph{$\mu$Serve}

\vspace{0pt}
\noindent \textbf{1. Scheduling Policies.}
Figure~\ref{fig:schedres} shows the P99 TTFT latency over time with different scheduling policies 
 driven by the production traces in Spitwise~\cite{splitwise} at 9 RPS.
%at high system load (9 RPS).
%Specifically,
We run \emph{S-LoRA} with its default FIFO scheduling policy~\cite{slora} and \emph{S-LoRA}
with the SJF scheduling policy from 
\emph{$\mu$Serve}~\cite{uServe}, as two state-of-the-art baselines.
Additionally, we run our proposed adapter-aware multi-queue scheduling policy, both without our caching mechanism (\emph{ChameleonNoCache}) and with it (\emph{\system}).

Both  \emph{S-LoRA} and \emph{S-LoRA}+SJF have large tail latencies that increase  over time due to the queuing bottlenecks. Their TTFT latencies amply violate the SLO.
%These designs have non-operational TTFT latencies.
With FIFO scheduling (\emph{S-LoRA}), the requests at the tail are short ones blocked by the earlier long 
ones,  while with SJF scheduling (\emph{S-LoRA+SJF}), the requests at the tail are long ones starved by the prioritization of short  requests.
Our proposed scheduling policy (\emph{ChameleonNoCache}) is very effective: it removes both HoL blocking effects and starvation, leading to much lower tail latencies.
Finally, by integrating our caching approach, the TTFT latency reduces further.

%The results in the figure are also aligned with observations from earlier experiments:
%to achieve the full performance potential
%the system needs to integrate both scheduling and caching approaches.
% , as \emph{\system{}} has the lowest tail latency.

\begin{figure}[t]
% \vspace{-2mm}
\centering
\includegraphics[width=0.9\columnwidth]{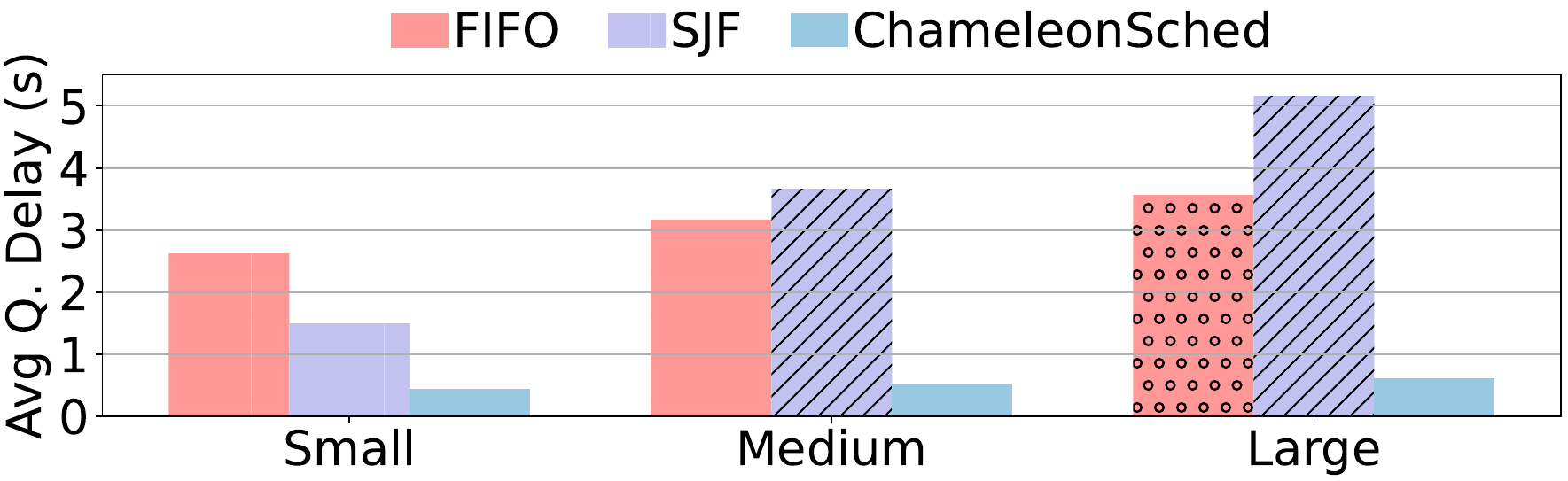}
\vspace{-5mm}
\caption{Average queuing time for each class of request
in  \emph{S-LoRA's FIFO},  \emph{SJF}, and the \emph{\system{}} Scheduler.}
\label{fig:queuedelays}
\vspace{-4mm}
\end{figure}

\vspace{0pt}
\noindent \textbf{2. Characterizing the Scheduling Policies.}
To understand why the \system{} Scheduler outperforms the other schedulers, we measure the time that requests spent waiting
in the queues before they are served. In Figure~\ref{fig:queuedelays} we plot the average queuing delays per request size category, as identified by \emph{\system{}} (small, medium, and large), and for the three scheduling policies, i.e. \emph{S-LoRA's FIFO}, \emph{SJF},
and the \emph{\system{}} Scheduler.
%that 
%each of the three 
%types of requests identified by \emph{\system{}}
%(short, medium, and long) being serviced. 
%Figure}~\ref{fig:queuedelays}\hl{  shows
%the average queuing time in %seconds for each class of request
%in \emph{S-LoRA}, \emph{$\mu$Serve}, and \emph{\system{}}.
We see that FIFO introduces relatively uniform absolute queuing delays. However, for small requests, queuing delays account for %$\approx$30\% 
28.6\% of their E2E latency. On the other hand, the SJF scheduler prioritizes small requests, creating long queuing delays for   large requests.
%($\approx$18\% of their E2E). 
Finally, the Chameleon scheduler substantially reduces queuing delays for all request types, bringing delays to below 8\% of the requests' E2E for all sizes.

\noindent \textbf{3. Caching Policies.}
We now compare different replacement policies for our proposed adapter cache.
Specifically, \emph{LRU} 
evicts from the cache the least recently used adapter.
\emph{FairShare} follows our proposed approach of considering the  adapter's recency, frequency, and size, but assigns 
the same weight to all   three knobs.
Finally, \emph{\system} uses our proposed algorithm, where the weights of the three knobs
are tuned based on our extensive profiling (Section~\ref{sec:cache}).

\begin{figure}[t]
% \vspace{-2mm}
\centering
\includegraphics[width=0.9\columnwidth]{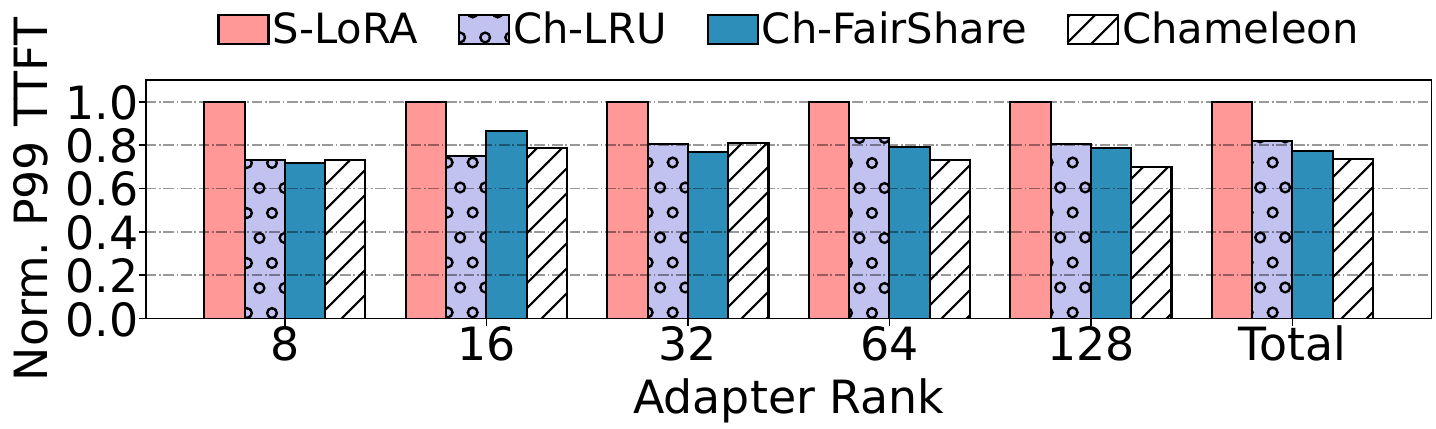}
\vspace{-4mm}
\caption{Normalized P99 TTFT latency for \emph{S-LoRA}, \emph{Chameleon-LRU}, \emph{Chameleon-FairShare}, and \emph{Chameleon}.}
\label{fig:res5a}
\vspace{-5mm}
\end{figure}

\vspace{0pt}

Figure~\ref{fig:res5a} shows the normalized P99 TTFT latency for requests of different adapter ranks 
at medium system load (8 RPS) for \emph{S-LoRA} (which does not have an adapter cache) and for 
\system{} with the three adapter
cache replacement policies described above. We see that 
\system{}'s proposed caching mechanism is very effective.
All the caching schemes reduce the P99 TTFT latency over \emph{S-LoRA} by a considerable amount for all adapter ranks.
Additionally, our proposed replacement policy further reduces the TTFT, especially for larger adapters.
For example, for requests with adapter rank 128, \emph{Chameleon} reduces the P99 TTFT latency over \emph{Ch-FairShare} by 12\%.
For the total trace,
\emph{Ch-LRU}, \emph{Ch-FairShare}, and \emph{Chameleon}
reduce the P99 TTFT latency over \emph{S-LoRA} by
18\%, 22\%, and 26\%, respectively.
%\hl{Add here comments for GDSF.}

% \noindent\textbf{5. GDSF Eviction Policy.}
\system{}'s eviction policy is based on cost and benefit estimations.
% , a strategy introduced by other
Prior work on software caches for objects with variable sizes proposed the Greedy Dual Size Frequency (GDSF) algorithm for web caching~\cite{GDSF}. GDSF uses $Score=$ $Frequency*Cost/Size +K$ to identify eviction candidates, where Cost is the overhead to load an object into the cache. 
\system{} applies this 
strategy to the new context of adapter caching, and proposes a new score formula for this use-case. GDSF's score is sub-optimal for a) the skewed access patterns of adapters, as it tends to cache only the most popular adapters and discards the rest, and b) the skewed rank popularity of adapters, as GDSF aggressively 
evicts larger adapters with moderate use frequency. %frequency may be evicted too aggressively,
%skewed 
% \hl{Instead, \system{} always promotes the caching of larger adapters. It correctly deems their cost of loading (i.e., the consumption of PCIe bandwidth), as more important than their cost of caching (i.e., the consumption of GPU memory). Its linearity also regulates the effect of frequency.}
%Figure~\ref{fig:gdsfandcoeff} \hl{
It can be shown that the
P99 TTFT for high load (9.5 RPS) and power-law adapter popularity for \emph{S-LoRA}
with the cache and eviction algorithm of   GDSF,
is substantially worse than that of Chameleon.

% \system{}  or 

\begin{figure}[t]
\vspace{0mm}
\centering
\includegraphics[width=0.9\columnwidth]{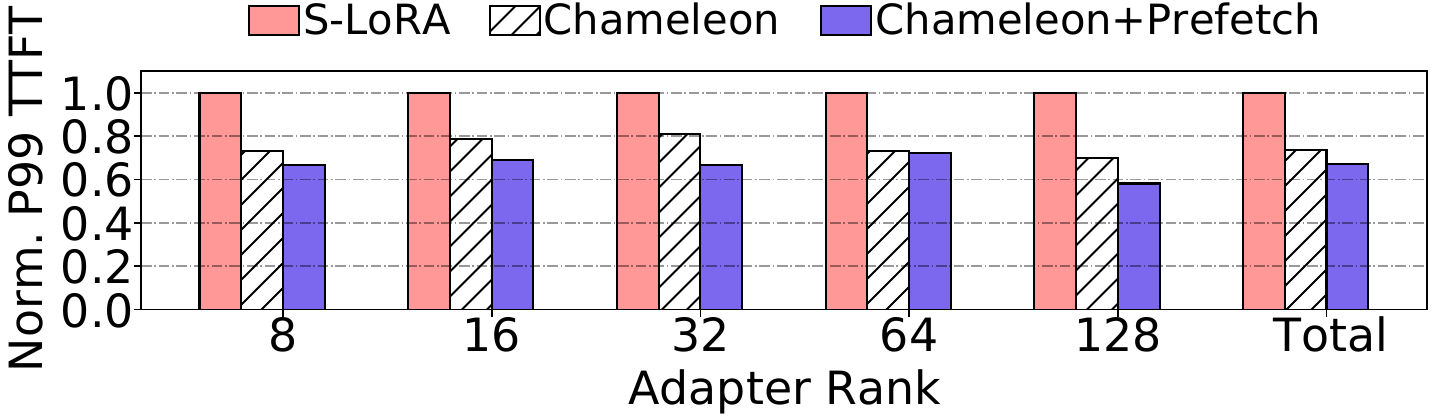}
\vspace{-4mm}
\caption{Normalized P99 TTFT latency for requests of different adapter ranks in \emph{S-LoRA}, \emph{Chameleon}, and \emph{Chameleon+Prefetch}.}
\label{fig:prefetch}
\vspace{-6mm}
\end{figure}

\noindent \textbf{4. Prefetching Mechanism.}
To reduce the latency of cache misses,
\system{} could use    prefetching.
It could predict which adapters are going to be used in the near future, and prefetch  them to the adapter cache ahead of time. To test this idea,
we have used a histogram-based technique to predict the future load of requests 
from~\cite{serverless_wild}. In this section, we show 
the potential benefits of using this prefetching. However, since 
the effectiveness of prefetching is highly dependent on the
prediction accuracy of  future loads, we do not include prefetching by default in  
 our  experiments in this paper.
%As this mechanism is highly dependent on the prediction accuracy of the future loads, we do not include it in any of our main experiments.

Figure~\ref{fig:prefetch} shows the normalized P99 TTFT latency for requests of different adapter ranks under medium load in three systems:
\emph{S-LoRA}, \emph{Chameleon}, and \emph{Chameleon+Prefetch}. We see that
prefetching can further reduce the TTFT latency of \emph{\system{}}. For the total trace, prefetching further reduces the P99 TTFT latency by 8.8\%.
As adapters are set to follow a uniform distribution for rank popularity
and a power-law distribution 
for adapter popularity within a rank,
their predictability is high. However, for other distributions, predictability may be lower.

\vspace{-1mm}
\subsection{Sensitivity Analysis}
\label{sec:newsensitivity}

To gain further insights into \system{}, we perform a sensitivity analysis of several of its
parameters.

\begin{comment}
%perform  sensitivity studies with respect to 
i) the accuracy of the output length predictor, 
ii) the  rank popularity of adapters,
iii) the total number of adapters in the system,
iv) additional traces,
% beyond those from,  %Splitwise}~\cite{splitwise}\hl{ 
v) another cache eviction
policy (GDSF), and
%the tuning of the coefficients in \system{}'s eviction policy and
vi) a nonlinear WRS formula. Due to space limits, we omit 
%all 
figures from this subsection.
\end{comment}

\label{sec:predictor}
\noindent\textbf{1. Impact of the Accuracy of the Output Length Predictor.}
Recall that the Chameleon Scheduler uses
an open-source BERT-based proxy model to predict a request’s
output length (Section~\ref{sec:scheduler}.1). We measure that our predictor has an average accuracy of about 80\%.
In this section, we examine the impact of artificially setting the predictor accuracy to 100\%, 80\%, and 60\%.
We consider two ways to compute the weighted
request size (WRS) (Section~\ref{sec:scheduler}.1): 
\emph{OutputOnly}, which uses  only the request output length (similar to~\cite{uServe}),
and \emph{\system{}}, which uses input and output length, and adapter size.

\begin{figure}[t]
\centering
\includegraphics[width=\columnwidth]{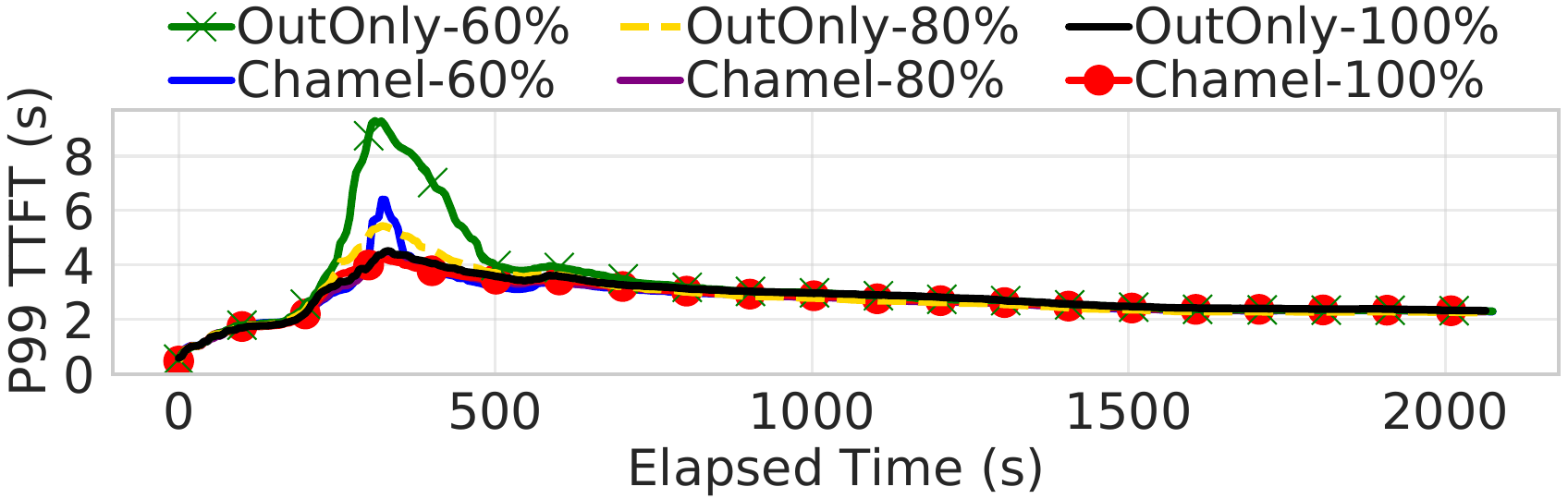}
\vspace{-6mm}
\caption{P99 TTFT latency over time for two configurations (\emph{OutputOnly} and \emph{\system{}}) under different output length predictor accuracies.}
\label{fig:accuracy}
\vspace{-5mm}
\end{figure}

Figure~\ref{fig:accuracy} shows the P99 TTFT latency for
\emph{OutputOnly} and \emph{\system{}} for the different output predictor accuracies as a function
of time. We see that the system is robust to predictor accuracy for most of the time.
However, during a load burst  
(at around 300s), the configurations with 60\% accuracy have high TTFT latency.
Also, the configuration that uses only the predicted output length (\emph{OutputOnly}) 
is more sensitive to the predictor accuracy than \emph{\system{}}.
%, \hl{as shown in the difference between \emph{OutputOnly-60\%} and \emph{\system{}-60\%}}. 
Finally, with a predictor of 80\% accuracy, \emph{\system{}} has approximately the same 
TTFT latency as with one of  100\% accuracy.

%the system has negligible performance loss compared to an ideal predictor (100\% accuracy).

\begin{comment}
\noindent\textbf{1. Impact of the Accuracy of the Output Length Predictor.}
We evaluate the efficiency of our scheduling policy under different accuracies of the output length predictor.
% CAMERA: 
\hl{Figure}~\ref{fig:accuracy} \hl{shows the P99 TTFT latency for} 
%We compare
two scheduling policies:
\emph{OutputOnly} considers only the request's output length~\cite{uServe},
% (similar to prior work~\cite{uServe}),
while \emph{\system{}} considers all request's properties (input and output length, and adapter size).
We run with 100\%, 80\%, and 60\% predictor accuracy.
% of the output length predictor.
In all other experiments, the accuracy is 80\%.

The system is robust to predictor accuracy for most of the time.
However, during load spikes 
\hl{(at around 300s)},
the systems with lower accuracy have high tail latency.
Importantly, the scheduler that uses only the predicted output length
% as knob to prioritize requests 
is more sensitive to the predictor accuracy than our proposed scheme, \hl{as shown in the difference between \emph{OutputOnly-60\%} and \emph{\system{}-60\%}}. We see that,
with the 80\% predictor accuracy, the system has negligible performance loss compared to an ideal predictor (100\% accuracy).
\end{comment}

\begin{figure}[t]
\centering
\includegraphics[width=\columnwidth]{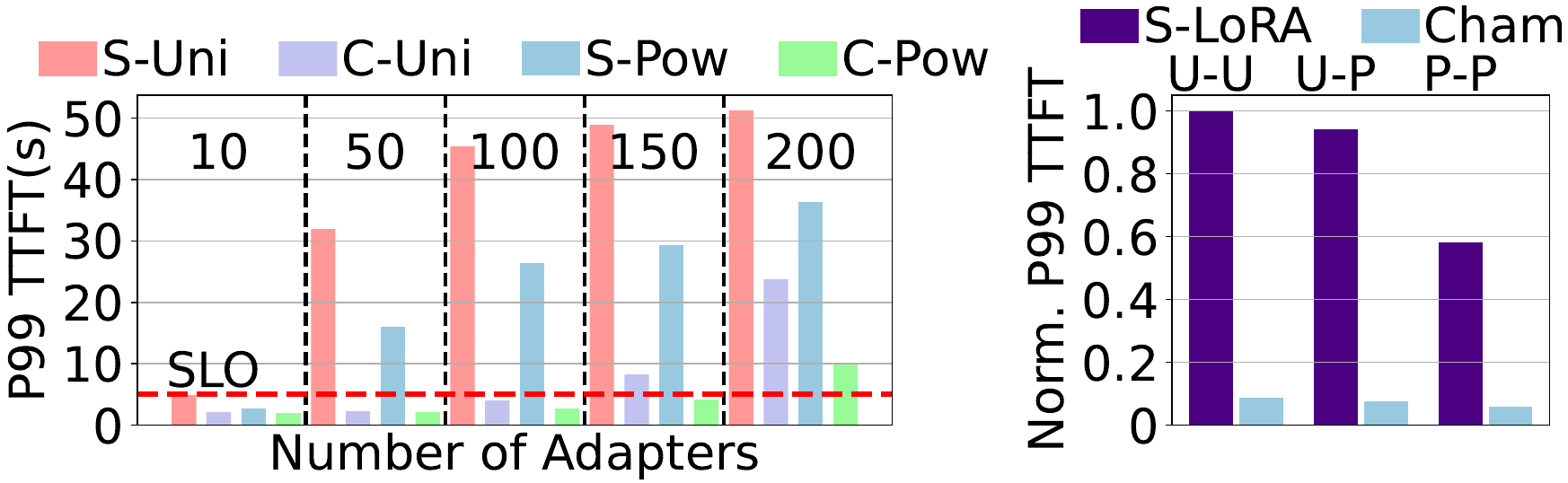}
\vspace{-6mm}
\caption{P99 TTFT latency sensitivity to  the total number of adapters (left) and to their distribution (right).}
\label{fig:skew}
\vspace{-3mm}
\end{figure}

%\subsubsection{\hl{Skewed popularity.}}
\label{sec:popularity}
\noindent\textbf{2. Impact of the distribution of adapter rank popularity and adapter popularity within a rank.}
By default, our experiments use a uniform distribution for adapter rank popularity and a
power-law distribution for adapter popularity within a rank
(Section~\ref{sec:methodology}). In this section,
we examine other distributions: 
i) uniform rank popularity and uniform adapter popularity within a rank \emph{(U-U)}, 
ii) uniform rank and power-law adapter popularity \emph{(U-P)}, 
%i.e., 30\% of the adapters are used by 50\% of the requests but they are uniformly distributed to ranks ranging from 8 to 128,
and
iii) power-law rank popularity and power-law adapter popularity \emph{(P-P)}. 
%i.e., 25\% of the adapters are used by 50\% of the requests and 80\% of them have rank lower or equal to 16.
%All skewed distributions use a power law distribution.
Figure~\ref{fig:skew}-right  shows the normalized
P99 TTFT 
latency with these distributions in \emph{S-LoRA} and  \emph{\system}.
We see that both \emph{S-LoRA} and  \emph{\system{}} perform best under the \emph{P-P} distribution,
as the cost of loading adapters and the queuing delays decrease. 
%However, \emph{S-LoRA} is more heavily affected, as it loads adapters more frequently, while 
\emph{\system{}}'s advanced caching and scheduling keep the P99 TTFT 
latency minimal for all distributions.

%To each request, we attach an adapter
%following a power-law distribution [49 ] for adapter popularity and
%a uniform distribution for rank popularity.
%{sec:methodology}

\begin{comment}
\begin{figure}[t]
\centering
\includegraphics[width=\columnwidth]{Source/figures/evaluation/nikoleta-distr}
\vspace{-6mm}
\caption{Sensitivity to the rank popularity.}
\label{fig:popularity}
\vspace{-4mm}
\end{figure}
\end{comment}

\begin{figure}[t]
\centering
\includegraphics[width=\columnwidth]{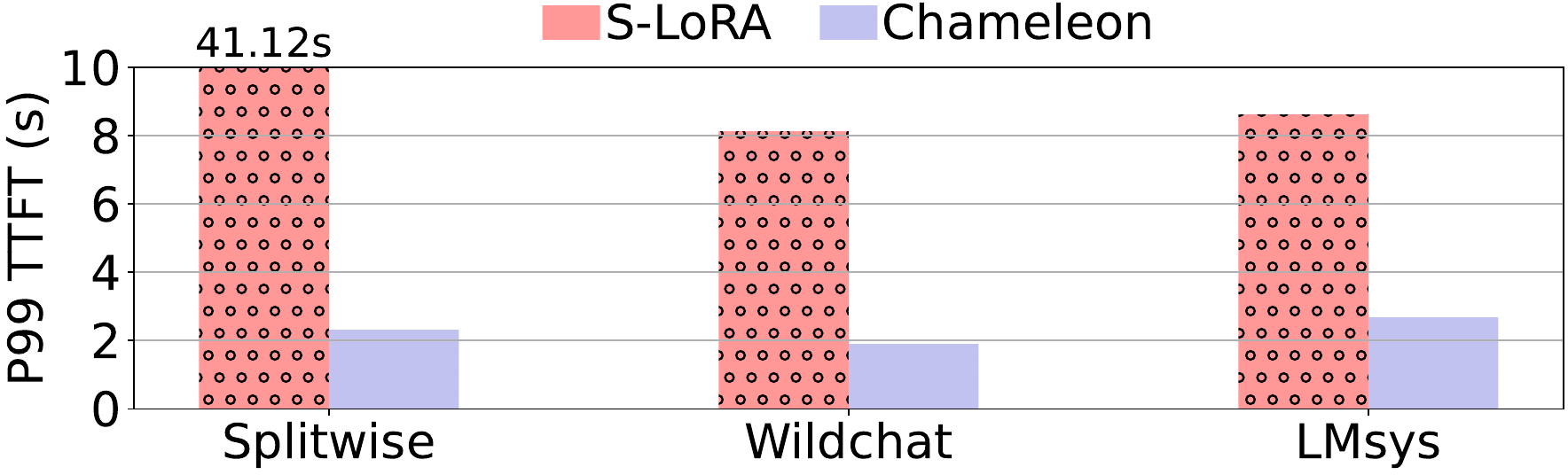}
\vspace{-6mm}
\caption{P99 TTFT latency for different traces. The SLOs for Splitwise, WildChat-1M, and LMSYS-Chat-1M are 5s, 3.3s, and 3.5s, respectively.}
\label{fig:inputoutputsensitivity}
\vspace{-5mm}
\end{figure}

%\subsubsection{\hl{Total number of adapters.}}
\label{sec:na}
\noindent\textbf{3. Impact of the total number of adapters.}
In our experiments, we have used a total number of adapters ($N_a$) equal to 100. 
In this section, we consider $N_a$ equal to 10, 50, 100, 150, and 200~\cite{slora,dlora-osdi}. We also consider both uniform and power-law distributions for the rank popularity. 
Figure~\ref{fig:skew}-left shows the P99 TTFT latency for \emph{S-LoRA} (S) and \emph{\system{}} (C) for uniform (Uni) and power-law (Pow) distributions. The load is 9.5 RPS and the SLO is 5s.
We see that
\emph{\system{}} keeps the TTFT under SLO for up to 100 adapters when using a  uniform 
distribution, and up to 150 when using a power-law distribution. In contrast, \emph{S-LoRA} can only meet SLO for either  distribution  for 10 adapters. As the number of adapters increases, \emph{\system{}} keeps the TTFT latency low because: i) its adapter cache minimizes the increasing overheads of adapter loading and ii) its scheduler reduces the increasing effect of HoL blocking.

%, as requests become more heterogeneous.

\noindent\textbf{4. Impact of Additional Traces.}
We now use different  traces beyond
those from Splitwise~\cite{splitwise} to evaluate \emph{\system{}}, without re-adjusting 
\emph{\system}'s tuned parameters---i.e., the coefficients in its cache eviction policy and WRS formula. We obtain traces from  two data-sets: WildChat-1M~\cite{wildchat} and LMSYS-Chat-1M~\cite{lmsys}.
In Figure~\ref{fig:inputoutputsensitivity}, we  plot the P99 TTFT latency for each  trace for 9.5 RPS. 
The SLOs for Splitwise, WildChat-1M, and LMSYS-Chat-1M are 5s, 3.3s, and 3.5s, respectively.
The new traces have generally smaller input and output lengths and thus their requests have shorter runtimes compared to Splitwise. In the figure,  we see that \emph{S-LoRA}  fails to meet the SLO under high load 
for all traces due to queuing. In contrast, \emph{\system{}} meets the SLOs for all traces, and 
reduces the TTFT latency in the new traces by about 4$\times$ over \emph{S-LoRA}.

%\subsubsection{\hl{GDSF.}}

% We see that, when the load spikes, the eviction policy of  \system{}   significantly outperforms that of   GDSF.}

%\hl{In the same Figure we also study \system{}'s sensitivity to the tuning of the}\todo{add content}.

\begin{figure}[t]
\centering
\includegraphics[width=0.9\columnwidth]{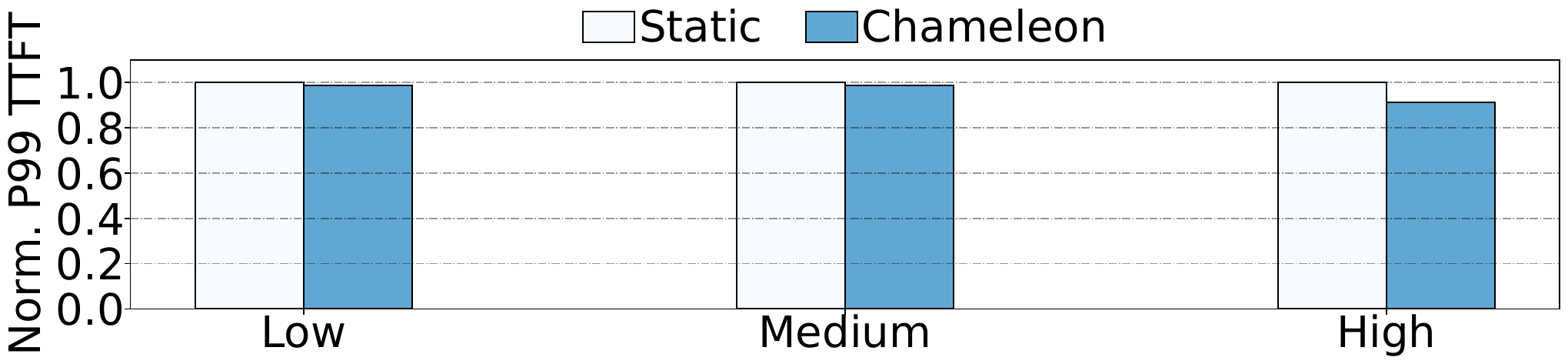}
\vspace{-4mm}
\caption{P99 TTFT latency for \emph{\system{}} normalized to a static scheme for different loads.}
%a \emph{static queues} approach for different load. %Latency is normalized to that of \emph{static queues} approach for a given load level (Low, Medium, High).
\label{fig:kmeans}
\vspace{-3mm}
\end{figure}

\noindent
\textbf{5. Impact of the Scheduling Queue Organization.}
\emph{\system{}} uses K-means clustering to  decide the number of scheduling queues and their cut-offs.
It then uses the equations in Section~\ref{sec:scheduler}.5 to assign resource quotas to queues. Further, it performs all  these
actions dynamically. In this section, we compare \emph{\system{}} to a static system that, knowing the
smallest and the largest size of requests, sets
the number of queues to 4, sets their ranges equally, and assigns the number of resource tokens
to each queue equally. We call the system {\em Static}. Figure~\ref{fig:kmeans} shows the normalized
P99 TTFT latency for {\em Static} and \emph{\system}. 
We see that, for low and medium load, the two configurations perform similarly. 
For high load, \emph{\system}'s design reduces the TTFT latency by 10\%.

\vspace{-1mm}
\subsection{Scalability Analysis}
\label{sub_scal}

To assess the scalability of  \system,
we run experiments with larger models (Llama-7B, Llama-13B, and Llama-30B) and 
with different memory capacities (24GB, 48GB, and 80GB).
In this section,
we run   the experiments on an A100 NVIDIA GPU that, by default, has 80GB of memory.
Given the available memory space, we use 500, 100, and 10 different adapters 
in the experiments with 7B, 13B, and 30B parameter models, respectively.

%with 7B, 13B, and 30B models.

\begin{figure}[t]
\centering
\includegraphics[width=0.9\columnwidth]{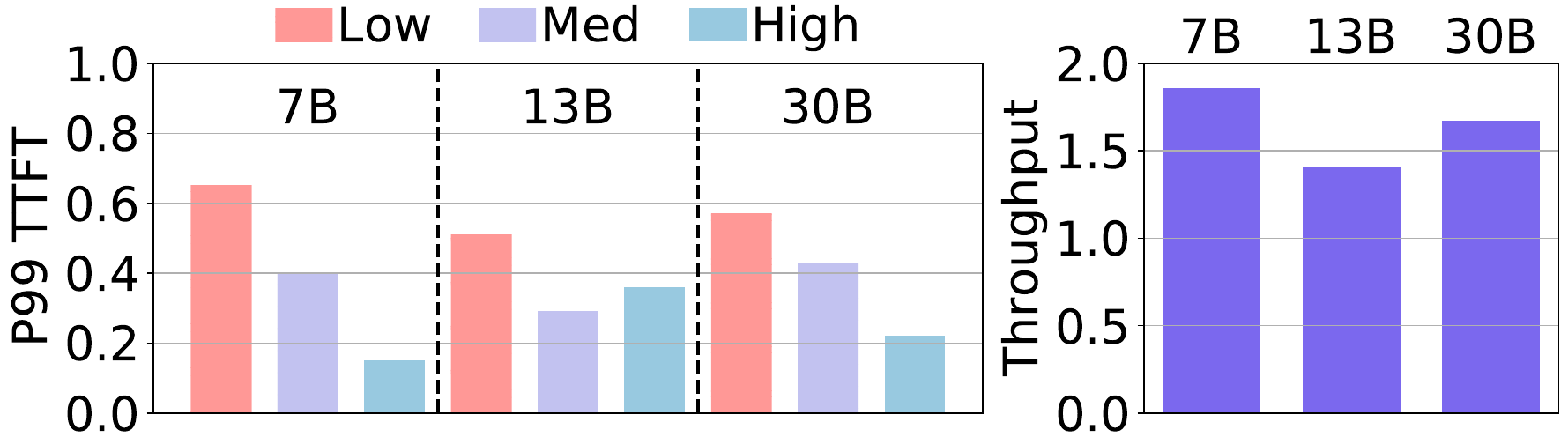}
\vspace{-4mm}
\caption{Normalized P99 TTFT latency (left) and throughput (right) of  \emph{\system{}} over \emph{S-LoRA} with different LLMs (Llama-7B, 13B, and 30B) and loads (Low, Medium, and High).}
\label{fig:scalability}
\vspace{-3mm}
\end{figure}

\begin{figure}[t]
\centering
\includegraphics[width=0.9\columnwidth]{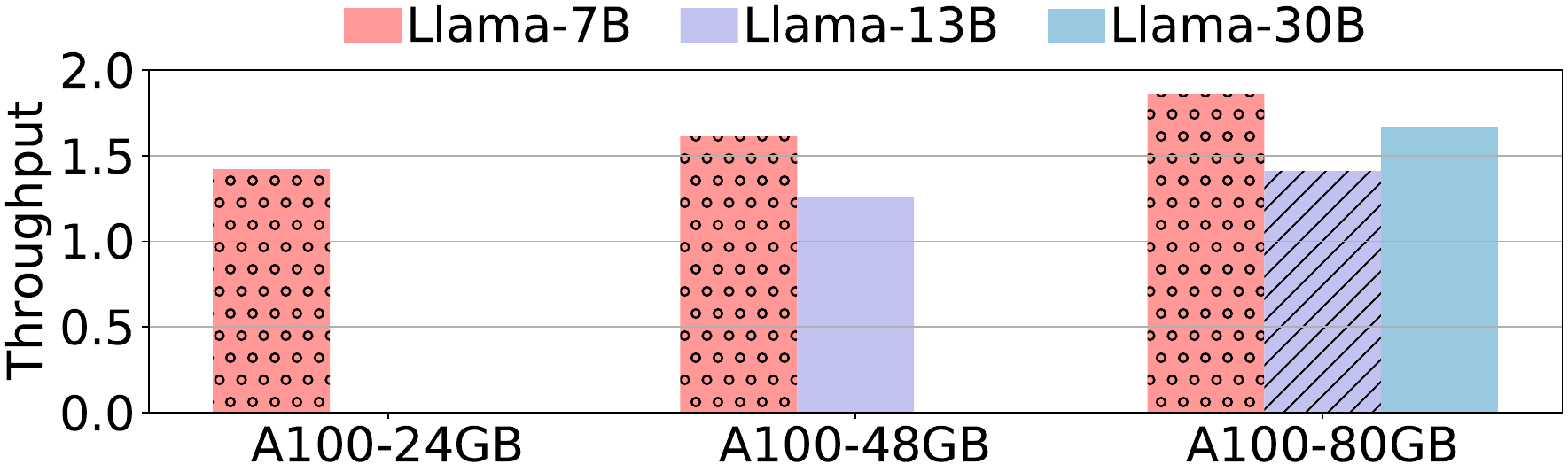}
\vspace{-4mm}
\caption{Normalized throughput of \emph{\system{}} over \emph{S-LoRA} with different 
GPU memory sizes (24GB, 48GB, and 80GB) and
LLMs sizes (Llama-7B, 13B, and 30B).}
% (24GB, 48GB, 80GB) on an A100 GPU.}
\label{fig:hardware}
\vspace{-4mm}
\end{figure}

%\vspace{0pt}
\noindent \textbf{1. Scalability with LLM size.}
In this section, we increase the size of the base LLM model and the load in the system.
Figure~\ref{fig:scalability}-left
shows the P99 TTFT latency 
of \emph{Chameleon} for  Llama-7B, 13B, and 30B, and for low, medium, and
high loads. The latency for 
a given model and load is normalized to \emph{S-LoRA}'s for the same model and load.
We see that \emph{Chameleon} always has a substantially lower TTFT latency than 
\emph{S-LoRA}.
\begin{comment}
As the model size increases, the 
request heterogeneity also increases
%heterogeneity across requests' execution times becomes more pronounced, 
i.e., %the difference in requests' sizes creates even
there are 
larger gaps in their execution times.
Thus, efficient scheduling provides more benefits.
On the other hand, as the load increases, the system with larger models uses more GPU memory, leaving less space available for caching.
%Therefore, a sweet-spot for \emph{\system{}}
%is medium load with the medium size LLM (Llama-13B).
\end{comment}
Overall, averaged across all loads,
\emph{\system{}}
reduces the P99 TTFT latency over \emph{S-LoRA}  by
60.0\%, 61.3\%, and 59.3\% for the Llama-7B, Llama-13B, and Llama-30B models, respectively.

Figure~\ref{fig:scalability}-right  shows the throughout of \emph{Chameleon} normalized to that of
\emph{S-LoRA} for different LLM sizes. We see that \emph{Chameleon} improves the 
throughout 
by
1.86$\times$, 1.41$\times$, and 1.67$\times$ for the
Llama-7B, Llama-13B, and Llama-30B models, respectively.

%shows the normalized throughput.
%of \emph{Chameleon} over \emph{S-LoRA} with different LLMs (Llama-7B, Llama-13B, and Llama-30B).
%\emph{\system{}} 
%also improves the throughout 
%over the baseline 
%by
%1.86$\times$, 1.41$\times$, and 1.67$\times$. 
%for Llama-7B, Llama-13B, and Llama-30B models, respectively.
%For larger models,
%\emph{\system{}}
%has less space to improve the throughput as such models are bounded by the memory capacity.
%For example, for both Llama-13B and Llama-30B, \emph{\system{}}
%increases the throughput 
%of \emph{S-LoRA} 
%from 2.5K and 1.5K tokens per second
%to 3.5K and 2.5K tokens per second, respectively.

%\vspace{0pt}
\noindent \textbf{2. Scalability with GPU memory size.}
In this section, we increase the GPU memory size in an A100 GPU.
%We consider performance scaling with different memory capacity.
Figure~\ref{fig:hardware} shows the normalized throughput of \emph{\system{}} over \emph{S-LoRA}
as we increase the
GPU memory size (24GB, 48GB, and 80GB) and the
LLM  size  (Llama-7B, 13B, and 30B).
Llama-30B
fits only in 80GB of memory,
Llama-13B fits in 48GB and 80GB of memory,
and Llama-7B fits in all memory configurations.
%As memory capacity increases,
%\emph{\system{}}
%improves its relative performance over the baseline to a larger extent.
We see that \emph{\system{}} is more effective at increasing the throughput over 
\emph{S-LoRA} as the amount of GPU memory increases. This is because
a larger memory creates more space for adapter caching.
For example, \emph{\system{}}
improves the throughput of Llama-7B  over \emph{S-LoRA}
by 1.4$\times$, 1.6$\times$, and 1.9$\times$
with 24GB, 48GB, and 80GB of GPU memory, respectively.

%\vspace{0pt}
\noindent \textbf{3. Scalability with GPU compute capability.}
This section compares the throughput improvements  of \emph{\system{}}
running on different hardware platforms with the same memory capacity.
We consider an A40 GPU with 48GB of memory and 100 adapters, and 
an A100 GPU with 48GB of memory and 500 adapters.
The first platform is discussed in Section~\ref{sec:results}.2 and Figure~\ref{fig:overall}.
\emph{\system{}} is shown to  improve the  throughput over \emph{S-LoRA} by 1.5$\times$. 
The second platform is discussed in Section~\ref{sub_scal}.2. As shown in the second bar
of Figure~\ref{fig:hardware}, 
\emph{\system{}} improves the  throughput over \emph{S-LoRA} by 1.6$\times$. Therefore,
\emph{\system{}}'s  improvement in throughput over \emph{S-LoRA} increases with the more powerful GPU,
even with more adapters.

\begin{comment}
We observe that 
\emph{\system{}} improves throughput over the baseline by 1.5$\times$ on an A40 GPU with 48GB of memory (Figure~\ref{fig:overall}),
%while Figure~\ref{fig:scalability} shows that 
%\emph{\system{}} improves that it 
and 1.9$\times$ on an A100 GPU with 48GB of memory (Figure~\ref{fig:scalability}).
%The reason 
%for larger performance savings on A100 over A40, 
% is that 
A100 
improves the request's execution time,
thus, adapter loading overheads substantially affect the request's critical path.
\end{comment}

\begin{comment}
    \noindent \textbf{3. Scalability to GPU compute capability.}
%In addition, 
%we analyze the performance scaling of \emph{\system{}}
%when 
We also compare Chame-leon's performance
%running 
on different hardware platforms with the same memory capacity.
We observe that 
\emph{\system{}} improves throughput over the baseline by 1.5$\times$ on an A40 GPU with 48GB of memory (Figure~\ref{fig:overall}),
%while Figure~\ref{fig:scalability} shows that 
%\emph{\system{}} improves that it 
and 1.9$\times$ on an A100 GPU with 48GB of memory (Figure~\ref{fig:scalability}).
%The reason 
%for larger performance savings on A100 over A40, 
% is that 
A100 
improves the request's execution time,
thus, adapter loading overheads substantially affect the request's critical path.
\end{comment}

\vspace{-2mm}
\subsection{Multi-GPU Experiments}
\label{sec:multigpu}

Finally, we evaluate \emph{\system{}} in a multi-GPU environment. We use the A100 server with 4 GPUs and employ tensor parallelism (TP) with  2  or 4 GPUs. We examine Low, Medium, and High request loads.
In this setup, the base LLaMA-7B model and the adapters are partitioned across the GPUs along tensor dimensions. \emph{\system{}}'s caching and scheduling mechanisms operate as in the single-GPU case. The cache is distributed across the GPUs, storing partitions of adapters, while scheduling continues to treat all GPUs as a single execution engine. No changes are made to the caching or scheduling policies to accommodate the multi-GPU setup.

\begin{figure}[t]
\centering
\includegraphics[width=0.9\columnwidth]{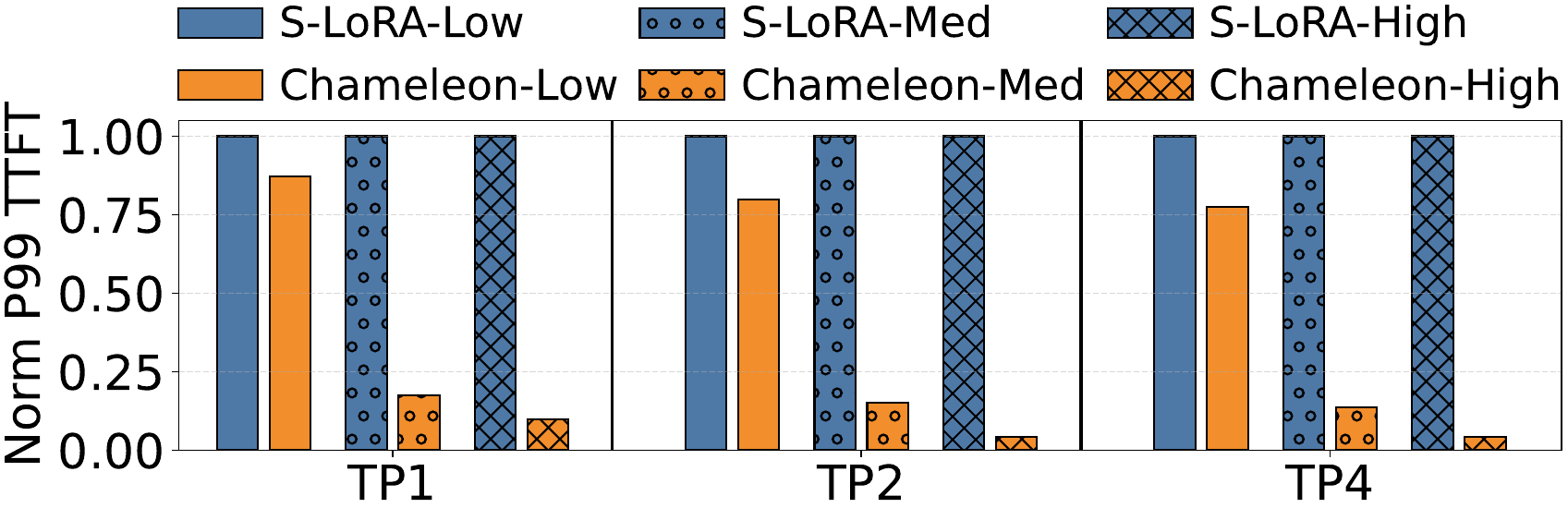}
\vspace{-4mm}
\caption{Normalized P99 TTFT latency for \emph{Chameleon} and \emph{S-LoRA} with different levels of 
tensor parallelism (TP1, TP2, and TP4), and request load  (Low, Medium, and High).}
\label{fig:multi-gpu}
\vspace{-5mm}
\end{figure}

Figure~\ref{fig:multi-gpu} compares the P99 TTFT latencies of \emph{\system{}} and \emph{S-LoRA}
for TP1, TP2, and TP4, and different request loads. The bars are normalized to \emph{S-LoRA} for the
specific level of parallelism and load. We see that
\emph{\system{}}  reduces the TTFT latency across all parallelism and load levels. The 
reduction widens with increasing parallelism. This is because, with more GPUs, the cost of loading adapters onto all participating GPUs becomes a bigger bottleneck in \emph{S-LoRA}. \emph{\system{}}'s ability to cache and reuse adapter fragments across GPUs helps it avoid this overhead and scale more efficiently. 
This effect gets accentuated at higher loads.
Overall, the gains of \emph{\system{}} are substantial: for TP4 and High load,  \emph{\system{}} 
reduces the P99 TTFT latency  by
95.8\% over \emph{S-LoRA}.

\vspace{-2mm}
\section{Related Work}
\label{sec:related}

%\vspace{2pt}
\noindent \textbf{LLM Inference Optimizations.}
Many  works proposed hardware \cite{splitwise,mecla,llmcompass,tender,fact,neupim,duplex,vga,flashllm,smartinfinity,piminference,lightransf}, algorithm~\cite{pregatedmoe,olive,vitality}
and system-level~\cite{specinfer,spotserve,vllm,orca,sarathi} optimizations for performance and energy-efficiency~\cite{polca,dynamollm,tapas} of LLM inference systems.
% Splitwise~\cite{splitwise}  splits the two phases of LLM inference requests (prefill and decode) on to separate machines while provisioning the resources for each phase independently.
% SpotServe~\cite{spotserve} is a distributed LLM serving system on preemptible instances in the cloud that dynamically reogranizes the inference cluster based on the current system state.
% In addition to optimizing the performance, researchers also explored the impact of LLM inference on power and energy efficiency in datacenters~\cite{polca,dynamollm}.
These works consider LLMs with only a base model and do not optimize for a multi-adapter LLM inference environment.
\system{} is orthogonal to such techniques and can be combined with them.
% for better performance or energy-efficiency.

% ISCA papers~\cite{splitwise,mecla,llmcompass,tender,pregatedmoe,olive,fact}

% ASPLOS papers~\cite{exegpt,neupim,specinfer,spotserve}

% MICRO papers~\cite{duplex,vga,flashllm}

% HPCA papers~\cite{smartinfinity,piminference,lightransf,vitality}

% Citations for the prior work~\cite{alpaserve, caraserve, dlora-osdi, dynamollm, falcon, fastserve, guideMetrics, llama2, llama3, llumnix, mixtral2, orca, punica, radford2019gpt, sarathi, slora, splitwise, vtc-scheduler, llmschedulingarxiv, uServe}

\myparagraph{LLM Inference with Parameter-Efficient Fine Tuning}
Since the adoption of parameter-efficient fine tuning techniques~\cite{lora,lorapro,liu2022ptuningv2prompttuning,lester2021powerscaleparameterefficientprompt},
% , such as LoRA adapters,
researchers have been working on optimizing the system stack for efficient LLM inference in multi-adapter environments~\cite{caraserve,dlora-osdi,punica,slora,pets}.
S-LoRA~\cite{slora} and Punica~\cite{punica} decouple the base model from task-specific adapters and fetch the required adapters on the fly from the host to the GPU memory.
% They store all adapters in the main memory and fetch the adapters used by the currently running queries to the GPU memory.
dLoRA~\cite{dlora-osdi} dynamically merges and unmerges adapters with the base model based on the current system state.
% Additionally, in a multi-node setup, dLoRA migrates requests and adapters between different worker replicas to maximize adapter re-use.
% \system{} builds on top of these works and proposes two new techniques: adapter caching and adapter-aware scheduling.
In the paper, we quantitatively compare \system{} to S-LoRA as the state-of-the-art baseline.

\myparagraph{LLM Inference Scheduling}
Many works explored scheduling policies for LLM inference serving
\cite{orca,fastserve,exegpt,uServe,llumnix,llmschedulingarxiv,vtc-scheduler}.
$\mu$Serve  \cite{uServe} 
and Learning to Rank~\cite{llmschedulingarxiv} 
reduce  the HoL blocking effects via %shortest-job-first
SJF
scheduling. 
% policy.
% In the paper, 
We quantitatively compare  to $\mu$Serve.
Based on input and output request lengths, % distribution,
 ExeGPT~\cite{exegpt} and DynamoLLM~\cite{dynamollm}
allocate resources (batch size and model parallelism) and schedule the requests for  minimal cost and energy consumption, respectively.
Llumnix~\cite{llumnix} reschedules the requests across worker replicas to improve load balance.
These works focus on   scheduling 
 LLM inference requests
in a multi-node environment, while using 
conventional 
iteration-level scheduling~\cite{orca} within a node.
% In contrast, 
\system{} redesigns the scheduling policy within a node, and can be combined with cluster-level schedulers.

%\cite{orca} Orca
%\cite{fastserve} FastServe
%\cite{exegpt} ExeGPT
%\cite{uServe} uServe
%\cite{llumnix} Llumnix
%\cite{llmschedulingarxiv} LLM scheduling
%\cite{vtc-scheduler} VTC
%\cite{uServe} $\mu$Serve and \cite{llmschedulingarxiv} Learning to Rank 

\myparagraph{General-Purpose Workload Scheduling} %\cite{qzilla,sita,sita2}
% \system{}'s multi-queue request scheduling is inspired by prior art on general-purpose workload scheduling in multi-server environments~\cite{qzilla,sita,sita2}.
Size-Interval Task Assignment (SITA)~\cite{sita,sita2} addresses head-of-line blocking by providing an “express-lane” for
short tasks.
% , protecting them from queuing behind rare, long ones.
Q-Zilla~\cite{qzilla} leverages  this idea and proposes a Server-Queue Decoupled Size-Interval Task Assignment
% (SQD-SITA) 
% SQD-SITA is a scheduling algorithm to minimize tail latency 
for highly diverse microservice invocations.
\system{} applies the algorithm to a new domain: multi-adapter LLM inference serving.
Moreover, SITA assumes perfect knowledge of task size, while Q-Zilla relies on request preemption.
On the other hand, 
\system{} uses a predictor for a request's output length (which is unknown ahead of time), and  does not use preemption due to its high cost in LLM inference environments~\cite{uServe,fastserve,vllm}.

% \vspace{-4ex}
\vspace{-4mm}
\section{Conclusion}
\label{sec:conclusion}

This paper presented Chameleon, an efficient LLM inference serving system for many-adapter environments. Chameleon introduces two new ideas: adapter caching and adapter-aware request
scheduling. 
% CAMERA:
Caching minimizes the overhead of loading the adapter weights on the request’s critical path, while scheduling alleviates   head-of-line blocking and starvation for requests with highly-diverse execution times. 
Under high loads, Chameleon reduces the P99 TTFT latency
by 80.7\% and the P50 TTFT latency by 48.1\% over a state-of-the-art-baseline, while improving the
throughput by 1.5$\times$.

\vspace{-3mm}
\begin{acks}

This work was supported  by NSF under grants CCF 2107470 and CCF 2316233; by ACE, one of the seven centers in JUMP 2.0, a Semiconductor Research Corporation (SRC) program sponsored by DARPA; by the IBM-Illinois Discovery Accelerator Institute; and by an Amazon 
%ML-Systems 
Fellowship funded by the UIUC AICE Center.

\end{acks}

\bibliographystyle{ACM-Reference-Format}
\bibliography{refs.bib}

%%%%%%% -- PAPER CONTENT ENDS -- %%%%%%%%

%\bibliographystyle{ACM-Reference-Format}
%\bibliography{sample-base}

\end{document}